\documentclass[%
 reprint,
superscriptaddress,
 amsmath,amssymb,
 aps,
prb,
]{revtex4-2}
\usepackage{comment}
\usepackage{graphicx}
\usepackage{dcolumn}
\usepackage{bm}
\usepackage{braket}

\usepackage{todonotes}

\begin{document}

\title{Superconductivity in the Fibonacci Chain}

\author{Ying Wang}
\email{wang116@usc.edu}
\affiliation{Department of Chemistry, University of Southern California}
\author{Gautam Rai}
\affiliation{Department of Physics and Astronomy, University of Southern California}
\affiliation{I. Institute of Theoretical Physics, Universit\"at Hamburg, Notkestraße 9-11, 22607 Hamburg, Germany}
\author{Chris Matsumura}
\affiliation{Department of Physics and Astronomy, University of Southern California}
\affiliation{Department of Physics, University of Washington, Seattle, WA 98195, USA}
\author{Anuradha Jagannathan}
\affiliation{
Laboratoire de Physique des Solides, Universit\'{e} Paris-Saclay, 91400 Orsay, France
}
\author{Stephan Haas}
\affiliation{Department of Physics and Astronomy, University of Southern California}

\begin{abstract}
Superconductivity was recently reported in several quasicrystalline systems. These are materials which are structurally ordered, but since they are not translationally invariant, the usual BCS theory does not apply. At the present time, the underlying mechanism and the properties of the superconducting phase are insufficiently understood. To gain a better understanding of quasiperiodic superconductors, we consider the attractive Hubbard model on the Fibonacci chain, and examine its low-temperature superconducting phase in detail using the Bogoliubov-de Gennes mean-field approach. We obtain superconducting solutions as a function of the parameters controlling the physical properties of the system: the strength of the Hubbard attraction $U$, the chemical potential $\mu$, and the strength of the modulation of the Fibonacci Hamiltonian, $w$. We find that there is a bulk transition at a critical temperature that obeys a power law in $U$. The local superconducting order parameter is self-similar both in real and perpendicular space. The local densities of states vary from site to site, however, the width of the superconducting gap is the same on all sites. The interplay between the Hubbard attraction and the intrinsic gaps of the Fibonacci chain results in a complex zero-temperature $\mu$-$U$ phase diagram with insulating domes surrounded by superconducting regions. Finally, we show that tuning $w$ from weak to strong quasicrystalline modulation gives rise to qualitatively different thermodynamic behaviors as could be observed by measuring the specific heat.
\end{abstract}

\maketitle


\section{\label{sec:intro} Introduction}
Superconductivity in a quasicrystal was first reported for an Al-Zn-Mg alloy in 2018 \cite{kamiyaDiscoverySuperconductivityQuasicrystal2018}, with a critical temperature of $0.05$~K. 
More recently, superconductivity has also been observed in the van der Waals layered
quasicrystal  $Ta_{1.6} Te$, with a bulk critical temperature $\sim 1K$ \cite{tokumoto2023superconductivity}, and in near-30 degree twisted bilayer graphene moir\'{e} quasicrystal \cite{Uri_2023}. These findings have raised questions regarding the nature of the superconducting instability and the structure of the Cooper pairs. Standard BCS theory for homogeneous systems does not, of course, apply for these systems, due to the absence of translational invariance.  While attempts have been made to employ the superposition of nearly degenerate eigenfunctions for constructing extended quasiperiodic Bloch wave functions in momentum space \cite{lesser2022emergence}, the task of providing an appropriate theoretical framework for describing interacting quasicrystals remains challenging. There have been a number of previous theoretical studies for two-dimensional models. For example, a real-space dynamical mean field  theory treatment of the negative-$U$ Hubbard model on the Penrose vertex model was used to study the spatial modulation of the superconducting order parameter \cite{sakaiSuperconductivityQuasiperiodicLattice2017}. Further studies have been carried out for models on the Penrose tiling \cite{takemori2020physical, hauckElectronicInstabilitiesPenrose2021, nagai2022intrinsic,liu2023unconventional} and the Ammann-Beenker tiling \cite{araujo2019conventional, fukushima2023supercurrent, nagai2022intrinsic}. Other numerical works on 2D models have investigated the superconducting state in the presence of a magnetic field \cite{sakai2019exotic} and topological superconductivity \cite{ghadimi2021topological, cao2020kohn, liu2023high}. The results from these studies  show that there are very complex variations in real space of the superconducting order parameter and of the local density of states. However, there is little understanding of these variations and how they depend on band filling, even on a qualitative level. 

To understand the systematics of spatial variations in such quasiperiodic systems, here we consider a paradigmatic one-dimensional model, i.e., the superconducting Fibonacci chain. We will examine in detail the dependence of the order parameter on the local environment, and interpret it in terms of approximate analytical solutions in a perturbative limit. The  model also allows us to analyze the dependence of the density of states, the specific heat, and the transition temperature on the parameters of the Hamiltonian, namely the hopping modulation strength, the chemical potential, and the interaction strength $U$.

\begin{figure}   
    \includegraphics[width = \columnwidth, trim = 20pt 5pt 60pt 50pt, clip]{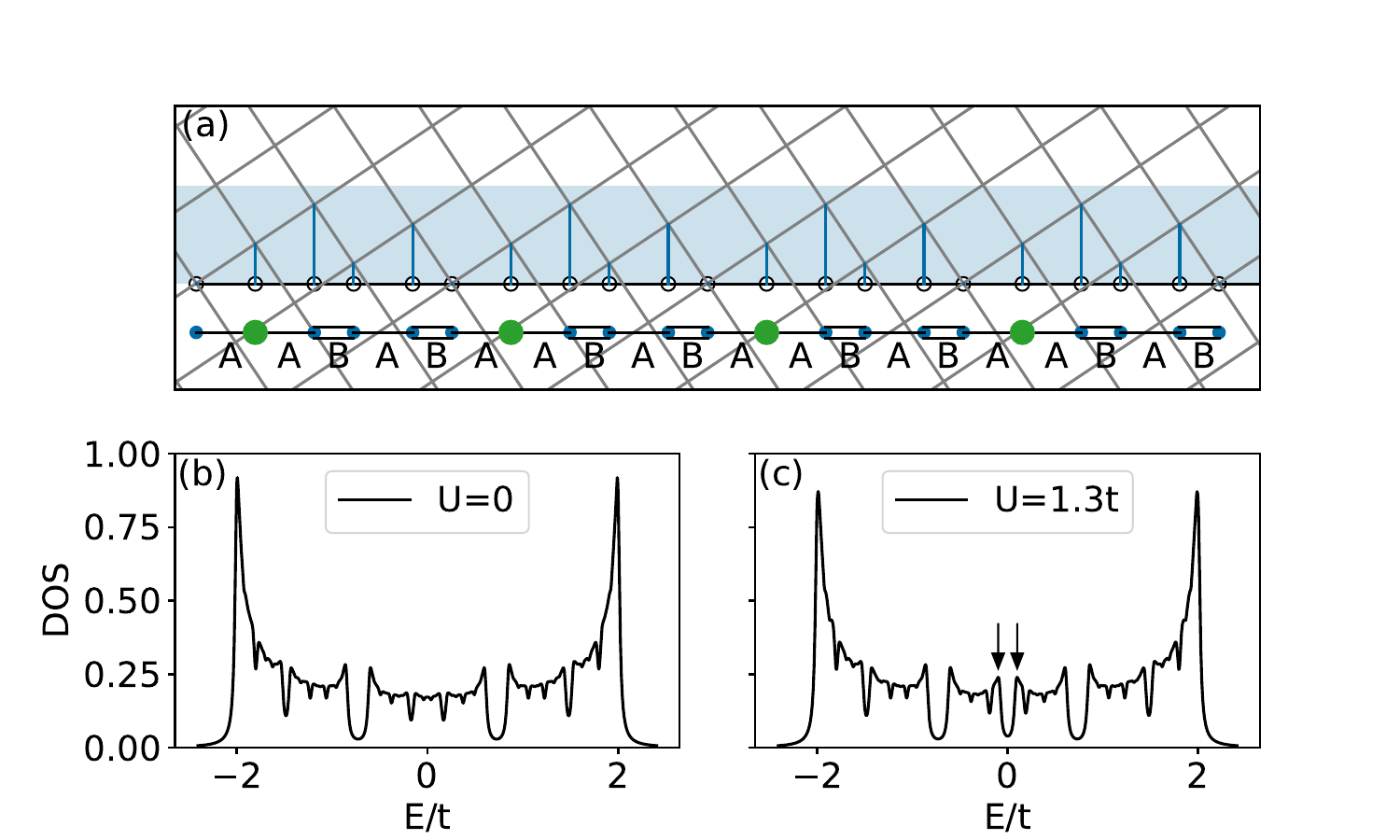}
    \caption{(a) Cut-and-project method to generate the Fibonacci chain: projecting selected sites of the 2D square lattice (within the blue strip) onto a line with slope equal to the golden ratio generates the Fibonacci chain. The chain linked by `A' and `B' hoppings is shown at the bottom of this figure. Atom sites (defined in Sec.~\ref{sec:model}) are marked by green in the Fibonacci chain. Global density of states in a 610-site approximant of the Fibonacci tight-binding chain (b) without and (c) with an attractive Hubbard $U$. When the attractive Hubbard interaction is introduced, a  superconducting gap appears in the density of states at the Fermi level, indicated by the arrows in (c). The modulation strength is $w=0.2$ in (b) and (c). The energy broadening width in density of states calculation (Lorentzian broadened) is $0.02t$.
    } 
    \label{cut-and-project and DOS}
\end{figure}

To this end,  we use the Bogoliubov-de Gennes framework to treat the negative-$U$ Hubbard model on the Fibonacci chain, assuming an inhomogeneous, i.e., site dependent $s$-wave superconducting order. We justify the use of mean field theory, by arguing that the 1D results can be carried over to a 3D extension of the Fibonacci chain, in which periodic 2D lattices are stacked along the third direction in a quasiperiodic way (See Appendix \ref{3D FC} ). Such Fibonacci superlattices, which have been studied both theoretically \cite{trabelsi2019photonic, trabelsi2019design, wu2012transmission,  woloszyn2012multifractal, korol2013energy} and experimentally \cite{todd1986synchrotron, cohn1988upper, zhu1997experimental}, justify the use of mean field theory which would otherwise not be meaningful for a true one-dimensional system. When the in-plane periodic hopping amplitudes are sufficiently weaker than the aperiodic hopping amplitudes in the out-of-plane direction, we find a close similarity between the results for the 1D and 3D systems.

To summarize the main results, we observe how local environments determine the local superconducting order parameter. Inherited from the Fibonacci hopping pattern, there is a self-similarity of the order parameter distribution in both real and perpendicular space. In addition, we study the variations of the local density of states (LDOS) in real space. While the details of the spectra, such as the heights of the coherence peaks, depend strongly on the local environment, the spectral gap which opens at the Fermi level is the same for all sites. This spectral gap is proportional to the critical temperature at which all local order parameters vanish, consistent with a bulk transition. We show that $T_c$ scales as a power law of $U$, where the power is a function of the modulation strength $w$. We calculate a zero-temperature phase diagram in the $\mu$-$U$ plane, which shows insulting domes surrounded by superconducting regimes. Finally, we show that the temperature dependence of the heat capacity shows qualitatively different behaviors in the weak, moderate and strong hopping modulation regimes.

\section{Model} \label{sec:model}

We consider an off-diagonal tight-binding model with a spatial modulation of the hopping integrals according to the Fibonacci series. This results in a binary hopping sequence generated by the infinite limit of the substitution rule $t_A \to t_A t_B$, $t_B \to t_A$ on an initial single hopping, $t_A$. The nearest-neighbor hopping integrals, $t_i$, take one of two values, $t_A$ or $t_B$, according to the Fibonacci sequence, where $i$ is the position index. In our numerical calculations, we use finite hopping sequences generated by applying the substitution rule a finite number of times denoted by $n$. These are the approximants of the Fibonacci chain---finite (periodic) structures that locally retain quasiperiodic character. The $n$th generation approximant contains a number of atoms equal to the Fibonacci number, $F_n = F_{n-1} + F_{n-2}$, with $F_0 = F_1 = 1$. We parameterize the strength of the quasiperiodic potential by the modulation strength $w = t_B - t_A$, with $t_B>t_A$. We fix the total bandwidth with the constraint that the average hopping $t = \frac{F_{n-1}t_A+F_{n-2}t_B}{F_n} = 1$. With this constraint, $t_A$ and $t_B$ are fully specified by $w$. The hopping ratio is given by $\rho = t_A/t_B = 1 - \frac{w\tau_n}{\tau_n + w}$, where $\tau_n = \frac{F_n}{F_{n-1}}$. The $\tau_n$ are called convergents of the golden ratio $\tau$, and $\tau_n\to\tau$ in the limit $n\to\infty$.

\begin{figure*}
    \includegraphics[width = 1\textwidth]{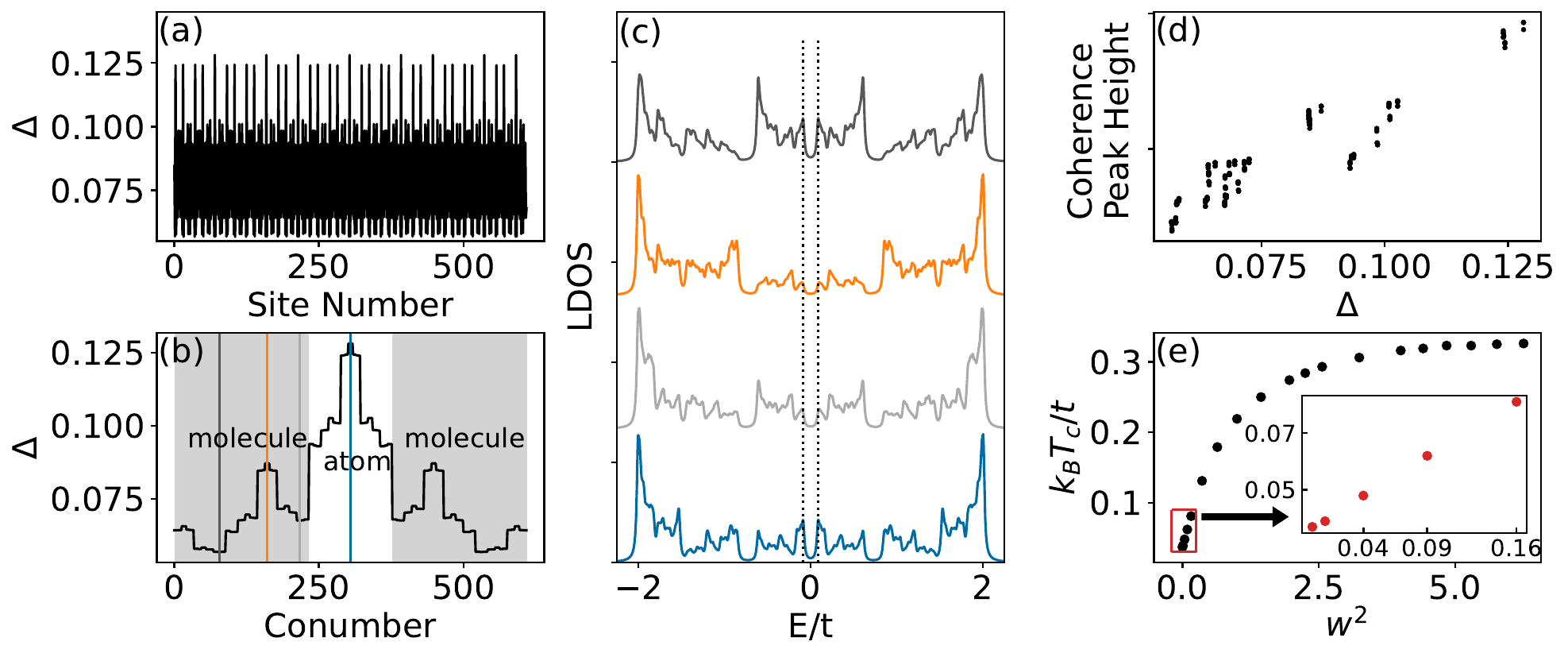}
    \caption{(a) Superconducting order parameter in real space and (b) in perpendicular space for a Fibonacci chain  with 610 sites at half-filling.  Panels (a)-(d) share all the same calculation parameters: the chain length is 610, the hopping modulation strength is $w=0.2$, and the BCS pairing strength is $U=1.3t$. (c) Local density of states (LDOS) at  four different sites marked in (b) with different representative values of the order parameter. The width of the central gap is the same in the LDOS for all sites. (d) Coherence peak height in the LDOS vs. the order parameter at the corresponding site. The order parameter at a given site is positively correlated with the height of the coherence peak in the local density of states for that site. (e) Dependence of the critical temperature on the square of the modulation strength for a chain with 610 sites and BCS pairing strength $U=1.3t$. The data points within the red frame in the main figure are magnified in the inset, which shows a linear scaling between critical temperature and the square of the modulation strength. In the inset of panel (e), the w values are 0.0,0.1,0.2,0.3,0.4 from left to right.}
    \label{fig:figure2}
\end{figure*}

Tight-binding models on the Fibonacci chain, the simplest quasicrystal, have been extensively studied for non-interacting electrons. These models have properties which are strikingly different from those of periodic systems, notably a fractal eigenvalue spectrum and states (see \cite{jagannathan2021fibonacci}). For a periodic approximant  of length $F_n$, there are $F_n$ distinct bands, and $F_{n}-1$ gaps. As $n\to\infty$, the gaps become dense, and the spectrum becomes singular continuous. As an example, in Fig.~\ref{cut-and-project and DOS} (b), we show the global density of states of a non-interacting 610-site approximant of the Fibonacci chain.

It was shown by Sire and Mosseri \cite{sireSpectrum1DQuasicrystals1989} that local environments can be easily classified in terms of the conumbers of sites. The conumber indexing is derived by considering the projection onto the perpendicular space in the cut-and-project scheme or model set method \cite{you1988global} of generating the Fibonacci chain. The conumber $c(i)\in[0,F_n-1]$ of a site with position index $i$ can be defined for a given Fibonacci approximant by~\cite{feingold1988universalities,sire1990excitation}
\begin{equation}\label{eq:conumber}
    c(i) = F_{n-1}i \mod F_n.
\end{equation}
The conumbering index classifies sites according to their local environment---sites  with similar local environments are closer to each other. In particular, the conumber indexing partitions sites into \emph{atom} or \emph{molecule} sites. There are $F_{n-3}$ sites that have a weak hopping $t_A$ in both directions. These are called atom sites and their conumber falls in the central window of the conumber sequence, $c(i)\in[F_{n-2},F_{n-1}]$. In Fig.~\ref{cut-and-project and DOS} (a), atom sites are marked by green dots. In contrast, the $2F_{n-2}$ molecule sites have a strong hopping $t_B$ in either the left or the right direction, and accordingly, their conumber lies in the right ($c(i)>F_{n-1}$) or the left ($c(i)<F_{n-2}$)window. This conumber representation was recently used in a study of spatial variations of the charge density, to show how these variations provide direct information on the topological indices of the chain \cite{raiBulkTopologicalSignatures2021}.

We now consider the superconducting phase induced by the attractive interaction $\hat{H}_I = \sum_i -\frac{U}{2}c^\dagger_{i\uparrow} c^\dagger_{i\downarrow} c_{i\downarrow} c_{i\uparrow}$. The Bogoliubov-de Gennes approximation \cite{de2018superconductivity} for this model results in introduction of two mean-field terms ---the Hartree shift $\epsilon_i^{H} = -U\langle c^\dag_{i\uparrow}c_{i\uparrow}\rangle$ and superconducting pairing amplitude $\Delta_i = U\langle c_{i\uparrow}c_{i\downarrow}\rangle$. The resulting mean-field Hamiltonian is
\begin{align}\label{Ham:effective}
     \hat{H} =& -\sum_{i\sigma} t_i (c^\dagger_{i+1\sigma}c_{i\sigma} + c^\dagger_{i\sigma}c_{i+1\sigma})\nonumber\\
     &+ \sum_{i\sigma}(\epsilon_{i} + \epsilon^{H}_{i} - \mu) c^\dagger_{i\sigma} c_{i\sigma}\nonumber \\
    &-U\sum_{i} \left(\Delta_i c^\dagger_{i\uparrow} c^\dagger_{i\downarrow}
    + \Delta^*_i c_{i\uparrow}c_{i\downarrow}\right),
\end{align}
where $c_{i\sigma}$ ($c^\dagger_{i\sigma}$) is the electron annihilation (creation) operator with spin $\sigma$ at site number $i$, $\epsilon_i$ is an on-site potential, and $\mu$ is the chemical potential.
 The second and third lines are the result of the mean-field treatment of the negative-$U$ Hubbard term. 
The mean-field quantities $\Delta_i$, representing the local superconducting order parameter (OP) and local Hartree energy $\epsilon_i^{HF}$ must satisfy self-consistency conditions,
\begin{align}\label{eq:selfCons}
    \Delta_i &= U\sum_n v^*_{in}u_{in}\left(1 - 2f(E_n, T)\right)\\
    \epsilon_i^{HF} &= U\sum_n|u_{in}|^2 f(E_n, T) + |v_{in}|^2(1 - f(E_n, T)),
\end{align}
where $E_n$ are the positive eigenvalues, and $(u_{in}, v_{in})$ are the eigenvectors of the Bogoliubov-de Gennes pseudo Hamiltonian \cite{de2018superconductivity} corresponding to \eqref{Ham:effective}. All of the results below are reported in units of $t$, the average hopping amplitude.

\section{Results}
\subsection{Order Parameter and critical temperature at half-filling} \label{basicpropoerties:halffilling}

 Figs.~\ref{cut-and-project and DOS}(b) and (c) show the density of states of the half-filled Fibonacci chain without and with the Hubbard attraction. In the absence of interactions, the density of states has gaps of varying widths throughout the spectrum. In the infinite system, the presence of a Hubbard attraction opens a superconducting gap at the Fermi level, with coherence peaks appearing on either side. In a finite system, this occurs as long as $U$ is larger than a threshold due to finite-size effects. This gap is distinct from the \emph{intrinsic} gaps of the non-interacting Fibonacci chain, which are also  present in (b). When $U$ is non-zero, and a gap opens, the singularities of the density of states of the non-interacting model are effectively averaged over an energy scale of order of the gap. The local superconducting order parameters $\Delta_i$ are shown in Fig.~\ref{fig:figure2}(a) for a 610-site approximant, at half-filling. The order parameter is position-dependent as the quasicrystal does not have translational symmetry. Its modulation follows the underlying quasiperiodic pattern. The complex real-space pattern, shown in Fig.~\ref{fig:figure2}(a), is self-similar, as shown in Appendix B. The complex spatial dependence is considerably simplified when transformed to conumber space, following Eq.~\eqref{eq:conumber}. Fig.~\ref{fig:figure2}(b) shows the same $\Delta_i$ as in (a), but with the indices permuted according to the conumber transformation. The site-by-site modulations observed in real space take a  layered structure, akin to a sequence of plateaus. Note that this is a self-similar fractal pattern, emerging also in conumber space. The self-similarity of OP distribution is explained in detail in Appendix \ref{selfsimilarity}. Furthermore, the conumber indexing allows us to connect the magnitude of the order parameter with the local neighborhood of the sites of the Fibonacci chain. We find that  the atom sites, with weak bonds to either side, have higher local order parameters than the molecule sites, with a strong bond to one side and a weak bond to the other side. This is a consequence of the fact that the eigenstates at half-filling have higher spectral weight on the atom sites than on the molecule sites \cite{niu1990spectral, piechon1995analytical}.

Even though the local order parameter is site-dependent, the superconducting gap width $\Delta_g$ in the local density of states is the same for all sites. This is shown by example of four representative sites with distinct neighbourhoods in Fig.~\ref{fig:figure2}~(c). The vertical dashed line marks the smallest energy with a non-zero value of the local density of states. This observation is a corollary of the fact that there is only one critical temperature although the order parameter varies locally, i.e., the entire Fibonacci chain becomes superconducting at the same critical temperature. This is consistent with the findings of \cite{araujo2019conventional} where the same is found to be true in the Ammann-Beenker tiling, a 2D quasicrystal, and in contrast to disordered systems, where strong disorder leads to the formation of superconducting islands \cite{ghosalInhomogeneousPairingHighly2001, ghosalRoleSpatialAmplitude1998}. The magnitude of the order parameter is  reflected by the strength of the coherence peaks. The positive correlation between the coherence peak heights and the magnitude of the order parameter is shown in the scatter plot in Fig.~\ref{fig:figure2}~(d). As can be expected in this mean field theory, the average order parameter $\Delta_{avg}$ scales as $\sqrt{T_c - T}$ close to and below the critical temperature (see Fig.~\ref{ameanOP-T} in Appendix~\ref{avgOPvsT}).

Fig.~\ref{fig:figure2}~(e) depicts the critical temperature $T_c$ plotted against the square of the modulation strength $w^2$. It shows that $T_c$ increases monotonically with $w$. This is a consequence of the fact that as $w$ is increased, the Fibonacci chain approaches the limit of disconnected atoms and molecules. In this limit, the intrinsic gaps of Fibonacci chain take up a higher proportion of the total band width. Since the bandwidth remains constant, the density of states in the ungapped regions must become higher as more states get squeezed into smaller energy intervals. This ultimately leads to a higher number of states being available to form the superconducting condensate around the Fermi level. The increase of critical temperature has been observed in other 1D   models \cite{fan2021enhanced, oliveira2023incommensurability}.
The functional form of the critical temperature as a function of the modulation strength can be understood in two opposite limits. In the weak modulation limit, $\rho\to 1, w\to 0$, the critical temperature scales quadratically in $w$ (inset of Fig.~\ref{fig:figure2}). This follows from treating the quasiperiodic modulation as a perturbation to the periodic system \cite{sireSpectrum1DQuasicrystals1989}, by which one can show that the intrinsic gaps of the Fibonacci chain grow linearly with $w$ in the perturbative regime \cite{raiBulkTopologicalSignatures2021}. In Appendix~\ref{app:sec:quadratic} we show how this translates to the BCS order parameter scaling quadratically in $w$. In the strong modulation limit, $\rho\to0, w\to$ $\sim$$2.6$, $T_c$ becomes essentially independent of $w$. In this limit, the Fibonacci chain decouples into a series of disconnected atoms (monomers) and molecules (dimers). In this limit, the density of states at half-filling takes the form of a delta function localized entirely on the atom sites, and therefore the critical temperature saturates to the value expected for a single-site Bogoliubov-de Gennes calculation, $k T_c \to U/4$ as $\rho\to 0$ (see Appendix \ref{Tc-larger-w}).

\begin{figure}
    \includegraphics[width = 1\columnwidth]{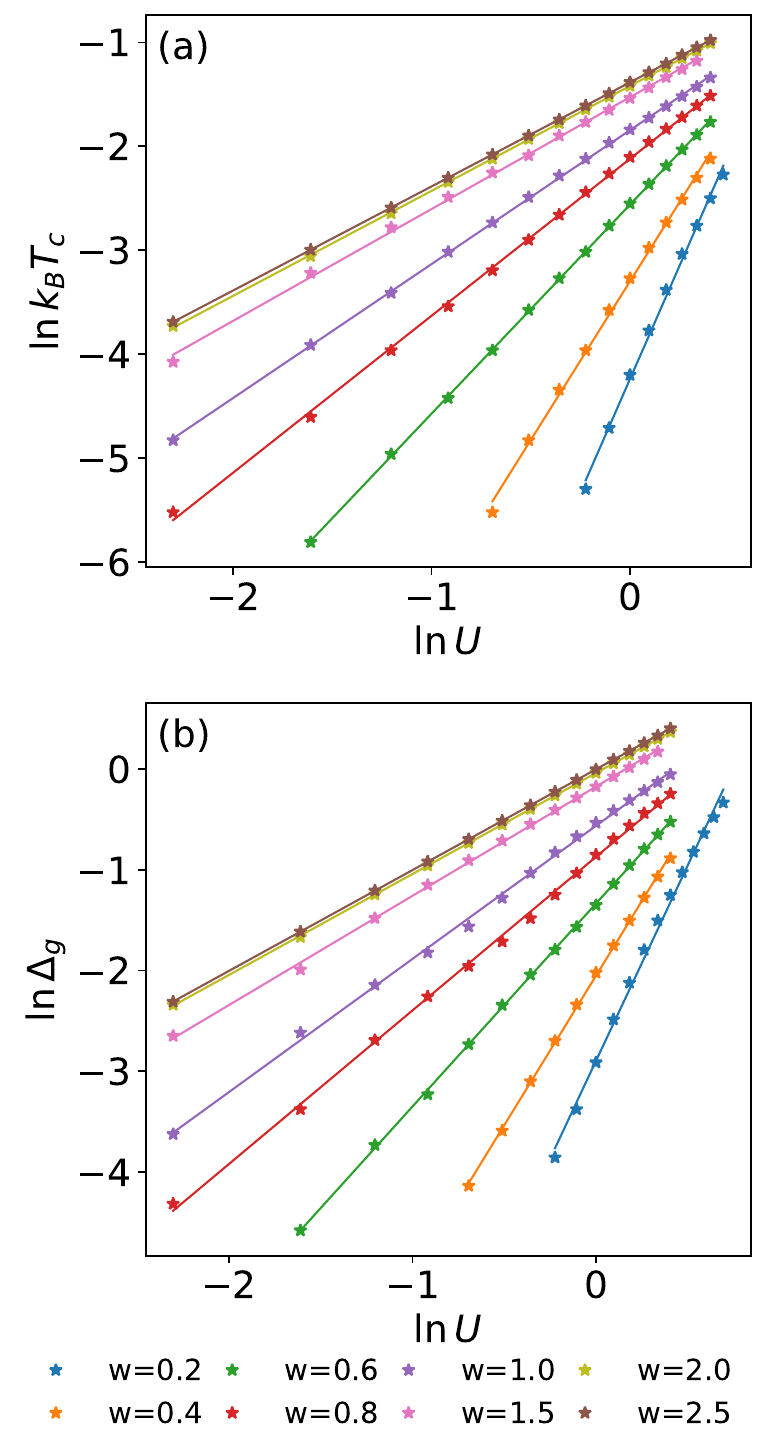}
    \caption{ (a) Scaling of the critical temperature  with the Hubbard attraction, shown on a log-log scale. Curves are shown for varying values of the modulation strength $w$ at half-filling.  The calculations are performed for an approximant  of 610 sites. These results indicate a power law dependence. (b) Scaling  of the superconducting gap width with the Hubbard attraction, also showing  a power-law scaling, with a $w$-dependent exponent.
}
    \label{fig:lnkBTc-lnU}
\end{figure}

\subsection{Scaling of the critical temperature and the superconducting gap width with $U$} \label{lnT_lnU and gapwidth_T}
In quasicrystals, the multifractality of the spectrum and of the eigenstates plays a crucial role in controlling the superconducting order parameter and critical temperature. The singularities of the spectrum offer a means to control these properties -- for example, to enhance the value of $T_c$ by tuning the chemical potential, $\mu$.

In this section we examine the scaling of $T_c$ and $\Delta_g$ as a function of the attractive interaction $U$ for different $w$ values. It has been shown \cite{noda2015bcs, noda2015magnetism} that when the density of states near the chemical potential scales as a power law, $\rho(E) \sim E^{-p}$, $T_c(U)$ should follow a power law $T_c \sim U^{1/\vert p \vert}$. In the non-interacting Fibonacci chain, the density of states is characterized by power law singularities throughout the spectrum. Let us consider the case of half-filling and the behavior of the density of states close to $E=0$. An exact solution was found for the scaling index at the center of the band \cite{kohmoto1987critical}. This solution predicts that $p>0$ increases  when the modulation strength is increased (see also \cite{zhong1995green}). Thus the power law scaling between $T_c$ and $U$ should depend on the modulation strength, as we show below.
    
In Fig.~\ref{fig:lnkBTc-lnU} (a), the dependence of $T_c$ on $U$ is shown for different values of $w$ in an approximant with 610 sites.  The critical temperatures and gap widths are global superconducting properties which are proportional to each other when $U$ is varied keeping other parameters constant, as shown in Fig.~\ref{gapwidthvsTc and STDEV} in Appendix~\ref{Tc-larger-w}.  Due to this proportionality, we also observe that the gap width $\Delta_g$ follows power-law scaling with respect to $U$, shown in Fig.~\ref{fig:lnkBTc-lnU} (b). For  fixed $U$, both $k_B T_c$ and $\Delta_g$ gradually increase and finally saturate as 
the modulation strength is enhanced toward the atomic limit. This scaling at half-filling illustrates that the quasiperiodic modulation strength is a sensitive tuning parameter to change the critical temperature and spectral gap width for a given value of interaction strength $U$. Other band fillings are observed to have similar behavior.

\begin{figure}
    \includegraphics[width = 1\columnwidth]{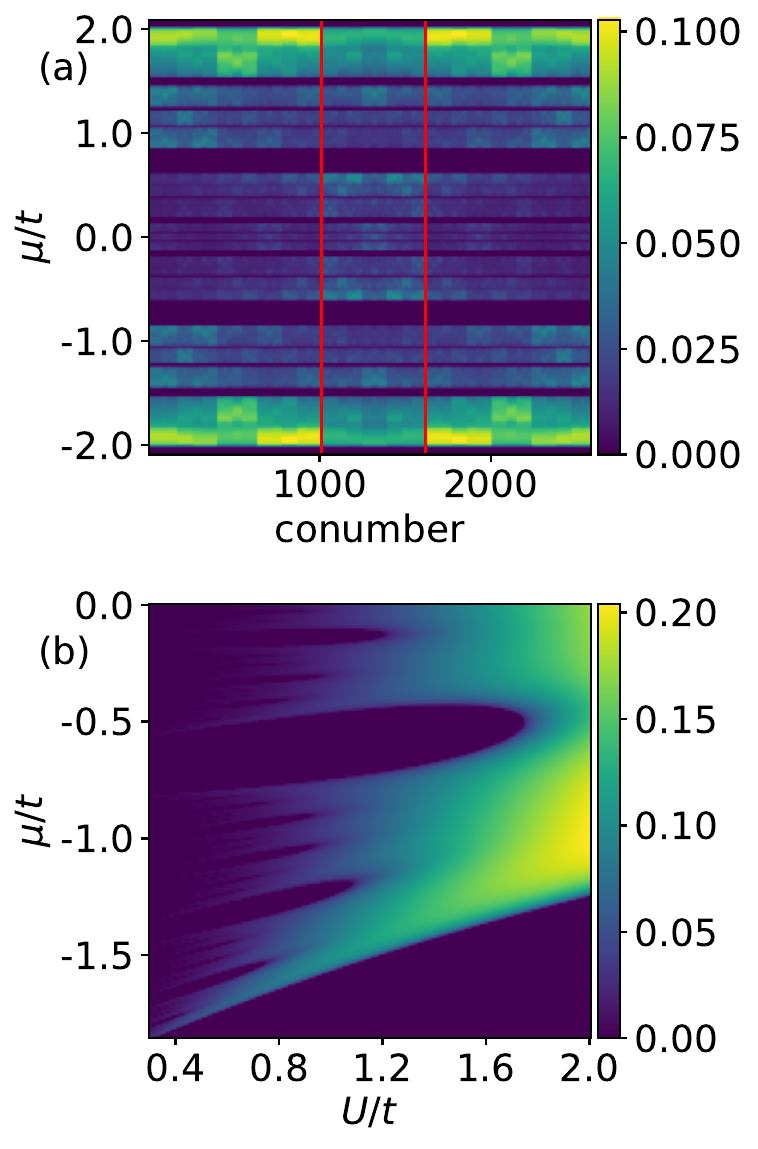} 
    \caption{(a) Local superconducting order parameter distribution in conumber space for a Hubbard attraction U=0.9t at various chemical potentials for a Fibonacci chain approximant of 2584 sites. (b) Phase Diagram: average superconducting order parameter as a function of the Hubbard attraction U and the chemical potential $\mu$. Insulating lobes occur where the superconducting order parameter vanishes (black regions). 
}
    \label{fig:phasediagram}
\end{figure}

\subsection{Superconducting order parameter as a function of band filling. $T=0$ phase diagram} \label{OP - band filling}
We now study the effects of varying the chemical potential $\mu$ in the Fibonacci chain. In Fig.~\ref{fig:phasediagram}(a), the color map shows the amplitude of the local order parameter in conumber space for a range of electron densities, parametrized by $\mu$ along the $y$-axis. In our convention, $\mu = 0$ represents the half-filled chain. If the Fermi level is tuned to a gapless part of the non-interacting spectrum, or if the intrinsic gap is not \emph{too large}, the system is superconducting below a finite $T_c$, and $\Delta_i$ is non-zero everywhere, with a  characteristic self-similar plateau structure akin to what is shown in Fig.~\ref{fig:figure2}(a), except that the relative values of the plateaus depends on the filling. On the coarse level, the relative magnitude of the local order parameter at a given site depends on the immediate local environment of the site. When the system is close to half-filling, atom sites (the sites between two red lines) have higher OP values; away from half filling, the molecule sites have higher OP values. The finer structure in Fig.~\ref{fig:phasediagram} is governed directly by the electron density around the Fermi level. Note the qualitative similarity between Fig.~\ref{fig:phasediagram}(a) and the local electron density map shown in Fig.~\ref{LDOS in conumber/energy} of \cite{mace2016fractal} (reproduced in Appendix \ref{LDOSinconumber/energy}). Fig.~\ref{fig:phasediagram}(a) also shows a self-similar pattern feature which is inherited from the spatial modulations of the Fibonacci chain couplings.

Upon varying the chemical potential $\mu$ and attraction $U$, we find the distribution of the average OP, shown in Fig.~\ref{fig:phasediagram} (b), which is closely related to the spectrum of the Fibonacci approximants. There are two distinct phases, i.e., superconducting and insulating,  that can be identified in this way, depending on whether the Fermi level lies in a region of finite density of states, or an intrinsic gap.

If the Fermi level is tuned to a gap, the system may be insulating or superconducting, based on a competition between the strength of the attraction $U$ and the width of the intrinsic gap.
A minimal model with this phenomenology is the BCS model for a semiconductor \cite{nozieresSemiconductorsSuperconductorsSimple1999, niroulaSpatialBCSBECCrossover2020}---a two band model with an attractive BCS term. In this model, when the Fermi level lies within a gap, there is a sharp transition from the superconducting to the insulating state as the Cooper pair binding energy crosses below the gap width. An analogous phenomenology is reflected in the Fibonacci chain, where the non-interacting spectrum contains a hierarchy of gaps. When the Fermi level is close to or within a particular gap, the low-energy physics can heuristically be approximated by the two-band model, as the fine structure in the density of states further away from the Fermi level is less relevant. Fig.~\ref{fig:phasediagram}(b) is a color map of the average order parameter in the chain as a function of $\mu$ and $U$. We see in Fig.~\ref{fig:phasediagram}(b), that as the BCS attraction is reduced, a hierarchy of insulating regions (with $\Delta_{avg}=0$) start to appear. Each of the insulating regions is associated with an \emph{intrinsic} gap of the Fibonacci chain. This gives rise to a complex phase diagram where at smaller values of $U$, a series of superconductor-insulator transitions occur as the chemical potential is raised or lowered. 

\begin{figure*}
    \centering
    \includegraphics[width = 1\textwidth]{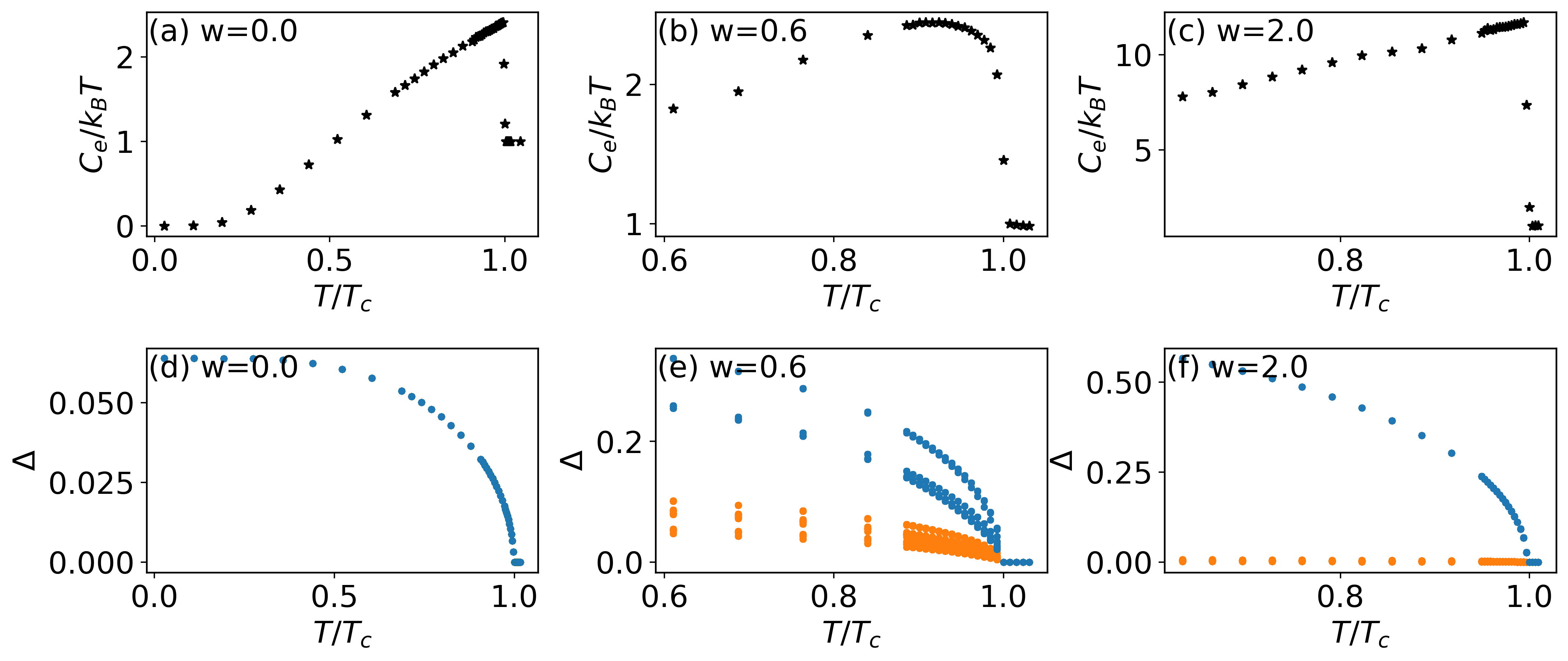}
    \caption{(a)-(c) Temperature dependence of the specific heat $C_e / k_B T$ for various modulation strengths $w$. For intermediate $w \in [0.5,0.8]$, a broadening of the peak is observed near the jump of specific heat. (d)-(f) Temperature dependence of the local superconducting order parameters. Blue (orange) dots represent the OP values of atom (molecule) sites. For intermediate $w$, we find a broad distribution of the local OP values, which lead to the thermal broadening of the peak in the specific heat. Results are shown for a Fibonacci chain approximant with 610 sites and $U = 1.3t$.}
    \label{Ce-T and Delta-T}
\end{figure*}

\subsection{Specific heat and temperature dependence of superconducting order parameter at half-filling}
We next consider the thermodynamics of superconducting Fibonacci chains at half-filling, as measured by the temperature dependence of the electronic specific heat,  
\begin{equation}
    C_e(T) = T \frac{dS(T)}{dT},
\end{equation}
where the entropy is defined by
\begin{equation}
    S(T) = 2\sum_{\alpha} \left[\ln{(1+e^{-\beta E_{\alpha}})}+ \frac{\beta E_{\alpha}}{e^{\beta E_{\alpha}}+1}\right].
\end{equation}

Here $E_{\alpha}$ are the eigenvalues of the Hamiltonian, and $\beta=1/k_BT$. Inspecting the numerical results shown in Figs.~\ref{Ce-T and Delta-T}, we observe three distinct regimes, depending on the modulation strength of the hopping parameter, $w$. The specific jump increases as $w$ increases. Similar to the analysis in \cite{takemori2020physical}, higher portion of eigenstates below $T_c$ leads to larger $C_e$. Close to the homogeneous limit ($w\rightarrow 0$), we recover the expected thermodynamic behavior of the uniform one-dimensional system, characterized by a sharply peaked specific heat that abruptly jumps to its normal state value at $T=T_c$, consistent with  mean field theory (Figs. ~\ref{Ce-T and Delta-T}(a) and (d)). In the opposite limit, $w > 1$, the system effectively breaks into its two constituents, i.e., atom sites  and two-site molecules, both of which are only weakly connected with their neighbors. For half filling, superconducting order is largest for the atom sites where the LDOS at the Fermi level is large (blue symbols in Fig.~\ref{Ce-T and Delta-T}(f)). In contrast, the LDOS at half-filling for molecule sites is very small, so that superconducting order is suppressed (orange symbols). The resulting distribution of the $\Delta_i$ is bi-modal, with one of the peaks close to zero, and the other peak at large values. In contrast, in the intermediate regime ($w \in [0.5,0.8]$) we observe a broad distribution of local superconducting OPs, reflecting the many different local environments, as shown in Fig. ~\ref{Ce-T and Delta-T}(e). This leads to a thermal broadening of the peak in the specific heat, as observed in Fig. ~\ref{Ce-T and Delta-T}(b). This broadening of the specific heat is noteworthy -- it is the signature of a regime in which the local SC order has a broad distribution of values. 

To summarize this section, the distribution of the order parameter has a strong effect on the behavior of the specific heat. Specifically, for narrow OP distributions ($w$ close to 0 or $w>1$), one finds the specific heat drops sharply at $T=T_c$. For broad distributions of the OP there is a rounding effect, that should also be observable in other thermodynamic quantities.

\section{Conclusions}

In conclusion, in this study we have examined local and global superconducting features of the quasicrystalline Fibonacci chain. 
Using a generalized, site-dependent mean-field approach, we have observed self-similar patterns of the local superconducting order parameter in real and perpendicular space. The local density of states vary from site to site, with large differences in the heights of the coherence peaks. However the width of the spectral gap remains the same throughout the chain. This is in keeping with the observation that there is a single transition temperature $T_c$ for this system, below which all of the local order parameters are non-zero. 

As is well-known, varying the hopping modulation results in large changes in the density of states of the Fibonacci chain. As a result, spectral gap width, and critical temperatures are strongly influenced by parameters such as the hopping modulation strength. They have a power law dependence on interaction strength $U$ with powers  that vary with modulation $w$.  For strongly modulated chains, the critical temperature can become much larger than that for the periodic chain. Enhanced critical temperatures have also been found in other models with fractal eigenstates such as the Aubry-André model \cite{fan2021enhanced}, generalized Aubry-André model \cite{oliveira2023incommensurability}, or in disordered models near the critical point \cite{feigel2010fractal}. However, although $T_c$ may be enhanced by increasing $w$, one should note that the distribution of local order parameters becomes highly inhomogeneous in this limit.  The temperature dependence of the specific heat is seen to depend strongly on the modulation strength of the hopping parameter, exhibiting a considerable smearing of the peak feature at intermediate strengths, where the local superconducting order parameter is distributed most widely.

The observed self-similarity in the superconducting order parameter suggests intriguing possibilities for tailoring and manipulating superconducting states in quasicrystalline materials. These insights could prove useful for the design of spatially inhomogeneous meta-materials with specifically tailored collective properties.  In conclusion, our results for the Fibonacci chain serve to illustrate the complexity of the new phases arising due to interactions combined with geometrical structure in quasicrystals. These systems can be expected to offer a rich platform for exploring novel superconducting phenomena.

\section{Acknowledgements}
The authors acknowledge the Center for Advanced Research Computing (CARC) at the University of Southern California for providing computing resources that have contributed to the research results reported within this publication. URL: https://carc.usc.edu. We also thank Dr. Yangyang Wan for providing  high performance computing resources. 

\appendix

\bibliographystyle{apsrev4-1}
\bibliography{SCinFC}

\begin{thebibliography}{44}%
\makeatletter
\providecommand \@ifxundefined [1]{%
 \@ifx{#1\undefined}
}%
\providecommand \@ifnum [1]{%
 \ifnum #1\expandafter \@firstoftwo
 \else \expandafter \@secondoftwo
 \fi
}%
\providecommand \@ifx [1]{%
 \ifx #1\expandafter \@firstoftwo
 \else \expandafter \@secondoftwo
 \fi
}%
\providecommand \natexlab [1]{#1}%
\providecommand \enquote  [1]{``#1''}%
\providecommand \bibnamefont  [1]{#1}%
\providecommand \bibfnamefont [1]{#1}%
\providecommand \citenamefont [1]{#1}%
\providecommand \href@noop [0]{\@secondoftwo}%
\providecommand \href [0]{\begingroup \@sanitize@url \@href}%
\providecommand \@href[1]{\@@startlink{#1}\@@href}%
\providecommand \@@href[1]{\endgroup#1\@@endlink}%
\providecommand \@sanitize@url [0]{\catcode `\\12\catcode `\$12\catcode
  `\&12\catcode `\#12\catcode `\^12\catcode `\_12\catcode `\%12\relax}%
\providecommand \@@startlink[1]{}%
\providecommand \@@endlink[0]{}%
\providecommand \url  [0]{\begingroup\@sanitize@url \@url }%
\providecommand \@url [1]{\endgroup\@href {#1}{\urlprefix }}%
\providecommand \urlprefix  [0]{URL }%
\providecommand \Eprint [0]{\href }%
\providecommand \doibase [0]{http://dx.doi.org/}%
\providecommand \selectlanguage [0]{\@gobble}%
\providecommand \bibinfo  [0]{\@secondoftwo}%
\providecommand \bibfield  [0]{\@secondoftwo}%
\providecommand \translation [1]{[#1]}%
\providecommand \BibitemOpen [0]{}%
\providecommand \bibitemStop [0]{}%
\providecommand \bibitemNoStop [0]{.\EOS\space}%
\providecommand \EOS [0]{\spacefactor3000\relax}%
\providecommand \BibitemShut  [1]{\csname bibitem#1\endcsname}%
\let\auto@bib@innerbib\@empty
\bibitem [{\citenamefont {Kamiya}\ \emph {et~al.}(2018)\citenamefont {Kamiya},
  \citenamefont {Takeuchi}, \citenamefont {Kabeya}, \citenamefont {Wada},
  \citenamefont {Ishimasa}, \citenamefont {Ochiai}, \citenamefont {Deguchi},
  \citenamefont {Imura},\ and\ \citenamefont
  {Sato}}]{kamiyaDiscoverySuperconductivityQuasicrystal2018}%
  \BibitemOpen
  \bibfield  {author} {\bibinfo {author} {\bibfnamefont {K.}~\bibnamefont
  {Kamiya}}, \bibinfo {author} {\bibfnamefont {T.}~\bibnamefont {Takeuchi}},
  \bibinfo {author} {\bibfnamefont {N.}~\bibnamefont {Kabeya}}, \bibinfo
  {author} {\bibfnamefont {N.}~\bibnamefont {Wada}}, \bibinfo {author}
  {\bibfnamefont {T.}~\bibnamefont {Ishimasa}}, \bibinfo {author}
  {\bibfnamefont {A.}~\bibnamefont {Ochiai}}, \bibinfo {author} {\bibfnamefont
  {K.}~\bibnamefont {Deguchi}}, \bibinfo {author} {\bibfnamefont
  {K.}~\bibnamefont {Imura}}, \ and\ \bibinfo {author} {\bibfnamefont {N.~K.}\
  \bibnamefont {Sato}},\ }\href {\doibase 10.1038/s41467-017-02667-x}
  {\bibfield  {journal} {\bibinfo  {journal} {Nature Communications}\ }\textbf
  {\bibinfo {volume} {9}},\ \bibinfo {pages} {154} (\bibinfo {year}
  {2018})}\BibitemShut {NoStop}%
\bibitem [{\citenamefont {Tokumoto}\ \emph {et~al.}(2023)\citenamefont
  {Tokumoto}, \citenamefont {Hamano}, \citenamefont {Nakagawa}, \citenamefont
  {Kamimura}, \citenamefont {Suzuki}, \citenamefont {Tamura},\ and\
  \citenamefont {Edagawa}}]{tokumoto2023superconductivity}%
  \BibitemOpen
  \bibfield  {author} {\bibinfo {author} {\bibfnamefont {Y.}~\bibnamefont
  {Tokumoto}}, \bibinfo {author} {\bibfnamefont {K.}~\bibnamefont {Hamano}},
  \bibinfo {author} {\bibfnamefont {S.}~\bibnamefont {Nakagawa}}, \bibinfo
  {author} {\bibfnamefont {Y.}~\bibnamefont {Kamimura}}, \bibinfo {author}
  {\bibfnamefont {S.}~\bibnamefont {Suzuki}}, \bibinfo {author} {\bibfnamefont
  {R.}~\bibnamefont {Tamura}}, \ and\ \bibinfo {author} {\bibfnamefont
  {K.}~\bibnamefont {Edagawa}},\ }\href@noop {} {\bibfield  {journal} {\bibinfo
   {journal} {arXiv preprint arXiv:2307.10679}\ } (\bibinfo {year}
  {2023})}\BibitemShut {NoStop}%
\bibitem [{\citenamefont {Uri}\ \emph {et~al.}(2023)\citenamefont {Uri},
  \citenamefont {de~la Barrera}, \citenamefont {Randeria}, \citenamefont
  {Rodan-Legrain}, \citenamefont {Devakul}, \citenamefont {Crowley},
  \citenamefont {Paul}, \citenamefont {Watanabe}, \citenamefont {Taniguchi},
  \citenamefont {Lifshitz}, \citenamefont {Fu}, \citenamefont {Ashoori},\ and\
  \citenamefont {Jarillo-Herrero}}]{Uri_2023}%
  \BibitemOpen
  \bibfield  {author} {\bibinfo {author} {\bibfnamefont {A.}~\bibnamefont
  {Uri}}, \bibinfo {author} {\bibfnamefont {S.~C.}\ \bibnamefont {de~la
  Barrera}}, \bibinfo {author} {\bibfnamefont {M.~T.}\ \bibnamefont
  {Randeria}}, \bibinfo {author} {\bibfnamefont {D.}~\bibnamefont
  {Rodan-Legrain}}, \bibinfo {author} {\bibfnamefont {T.}~\bibnamefont
  {Devakul}}, \bibinfo {author} {\bibfnamefont {P.~J.~D.}\ \bibnamefont
  {Crowley}}, \bibinfo {author} {\bibfnamefont {N.}~\bibnamefont {Paul}},
  \bibinfo {author} {\bibfnamefont {K.}~\bibnamefont {Watanabe}}, \bibinfo
  {author} {\bibfnamefont {T.}~\bibnamefont {Taniguchi}}, \bibinfo {author}
  {\bibfnamefont {R.}~\bibnamefont {Lifshitz}}, \bibinfo {author}
  {\bibfnamefont {L.}~\bibnamefont {Fu}}, \bibinfo {author} {\bibfnamefont
  {R.~C.}\ \bibnamefont {Ashoori}}, \ and\ \bibinfo {author} {\bibfnamefont
  {P.}~\bibnamefont {Jarillo-Herrero}},\ }\href {\doibase
  10.1038/s41586-023-06294-z} {\bibfield  {journal} {\bibinfo  {journal}
  {Nature}\ }\textbf {\bibinfo {volume} {620}},\ \bibinfo {pages} {762–767}
  (\bibinfo {year} {2023})}\BibitemShut {NoStop}%
\bibitem [{\citenamefont {Lesser}\ and\ \citenamefont
  {Lifshitz}(2022)}]{lesser2022emergence}%
  \BibitemOpen
  \bibfield  {author} {\bibinfo {author} {\bibfnamefont {O.}~\bibnamefont
  {Lesser}}\ and\ \bibinfo {author} {\bibfnamefont {R.}~\bibnamefont
  {Lifshitz}},\ }\href@noop {} {\bibfield  {journal} {\bibinfo  {journal}
  {Physical Review Research}\ }\textbf {\bibinfo {volume} {4}},\ \bibinfo
  {pages} {013226} (\bibinfo {year} {2022})}\BibitemShut {NoStop}%
\bibitem [{\citenamefont {Sakai}\ \emph {et~al.}(2017)\citenamefont {Sakai},
  \citenamefont {Takemori}, \citenamefont {Koga},\ and\ \citenamefont
  {Arita}}]{sakaiSuperconductivityQuasiperiodicLattice2017}%
  \BibitemOpen
  \bibfield  {author} {\bibinfo {author} {\bibfnamefont {S.}~\bibnamefont
  {Sakai}}, \bibinfo {author} {\bibfnamefont {N.}~\bibnamefont {Takemori}},
  \bibinfo {author} {\bibfnamefont {A.}~\bibnamefont {Koga}}, \ and\ \bibinfo
  {author} {\bibfnamefont {R.}~\bibnamefont {Arita}},\ }\href {\doibase
  10.1103/PhysRevB.95.024509} {\bibfield  {journal} {\bibinfo  {journal}
  {Physical Review B}\ }\textbf {\bibinfo {volume} {95}},\ \bibinfo {pages}
  {024509} (\bibinfo {year} {2017})}\BibitemShut {NoStop}%
\bibitem [{\citenamefont {Takemori}\ \emph {et~al.}(2020)\citenamefont
  {Takemori}, \citenamefont {Arita},\ and\ \citenamefont
  {Sakai}}]{takemori2020physical}%
  \BibitemOpen
  \bibfield  {author} {\bibinfo {author} {\bibfnamefont {N.}~\bibnamefont
  {Takemori}}, \bibinfo {author} {\bibfnamefont {R.}~\bibnamefont {Arita}}, \
  and\ \bibinfo {author} {\bibfnamefont {S.}~\bibnamefont {Sakai}},\
  }\href@noop {} {\bibfield  {journal} {\bibinfo  {journal} {Physical Review
  B}\ }\textbf {\bibinfo {volume} {102}},\ \bibinfo {pages} {115108} (\bibinfo
  {year} {2020})}\BibitemShut {NoStop}%
\bibitem [{\citenamefont {Hauck}\ \emph {et~al.}(2021)\citenamefont {Hauck},
  \citenamefont {Honerkamp}, \citenamefont {Achilles},\ and\ \citenamefont
  {Kennes}}]{hauckElectronicInstabilitiesPenrose2021}%
  \BibitemOpen
  \bibfield  {author} {\bibinfo {author} {\bibfnamefont {J.~B.}\ \bibnamefont
  {Hauck}}, \bibinfo {author} {\bibfnamefont {C.}~\bibnamefont {Honerkamp}},
  \bibinfo {author} {\bibfnamefont {S.}~\bibnamefont {Achilles}}, \ and\
  \bibinfo {author} {\bibfnamefont {D.~M.}\ \bibnamefont {Kennes}},\
  }\href@noop {} {\bibfield  {journal} {\bibinfo  {journal} {Physical Review
  Research}\ }\textbf {\bibinfo {volume} {3}},\ \bibinfo {pages} {023180}
  (\bibinfo {year} {2021})},\ \bibinfo {note} {publisher: APS}\BibitemShut
  {NoStop}%
\bibitem [{\citenamefont {Nagai}(2022)}]{nagai2022intrinsic}%
  \BibitemOpen
  \bibfield  {author} {\bibinfo {author} {\bibfnamefont {Y.}~\bibnamefont
  {Nagai}},\ }\href@noop {} {\bibfield  {journal} {\bibinfo  {journal}
  {Physical Review B}\ }\textbf {\bibinfo {volume} {106}},\ \bibinfo {pages}
  {064506} (\bibinfo {year} {2022})}\BibitemShut {NoStop}%
\bibitem [{\citenamefont {Liu}\ \emph {et~al.}(2023{\natexlab{a}})\citenamefont
  {Liu}, \citenamefont {Shao}, \citenamefont {Cao},\ and\ \citenamefont
  {Yang}}]{liu2023unconventional}%
  \BibitemOpen
  \bibfield  {author} {\bibinfo {author} {\bibfnamefont {Y.-B.}\ \bibnamefont
  {Liu}}, \bibinfo {author} {\bibfnamefont {Z.-Y.}\ \bibnamefont {Shao}},
  \bibinfo {author} {\bibfnamefont {Y.}~\bibnamefont {Cao}}, \ and\ \bibinfo
  {author} {\bibfnamefont {F.}~\bibnamefont {Yang}},\ }\href@noop {} {\bibfield
   {journal} {\bibinfo  {journal} {arXiv preprint arXiv:2306.12641}\ }
  (\bibinfo {year} {2023}{\natexlab{a}})}\BibitemShut {NoStop}%
\bibitem [{\citenamefont {Ara{\'u}jo}\ and\ \citenamefont
  {Andrade}(2019)}]{araujo2019conventional}%
  \BibitemOpen
  \bibfield  {author} {\bibinfo {author} {\bibfnamefont {R.~N.}\ \bibnamefont
  {Ara{\'u}jo}}\ and\ \bibinfo {author} {\bibfnamefont {E.~C.}\ \bibnamefont
  {Andrade}},\ }\href@noop {} {\bibfield  {journal} {\bibinfo  {journal}
  {Physical Review B}\ }\textbf {\bibinfo {volume} {100}},\ \bibinfo {pages}
  {014510} (\bibinfo {year} {2019})}\BibitemShut {NoStop}%
\bibitem [{\citenamefont {Fukushima}\ \emph {et~al.}(2023)\citenamefont
  {Fukushima}, \citenamefont {Takemori}, \citenamefont {Sakai}, \citenamefont
  {Ichioka},\ and\ \citenamefont {Jagannathan}}]{fukushima2023supercurrent}%
  \BibitemOpen
  \bibfield  {author} {\bibinfo {author} {\bibfnamefont {T.}~\bibnamefont
  {Fukushima}}, \bibinfo {author} {\bibfnamefont {N.}~\bibnamefont {Takemori}},
  \bibinfo {author} {\bibfnamefont {S.}~\bibnamefont {Sakai}}, \bibinfo
  {author} {\bibfnamefont {M.}~\bibnamefont {Ichioka}}, \ and\ \bibinfo
  {author} {\bibfnamefont {A.}~\bibnamefont {Jagannathan}},\ }in\ \href@noop {}
  {\emph {\bibinfo {booktitle} {Journal of Physics: Conference Series}}},\
  Vol.\ \bibinfo {volume} {2461}\ (\bibinfo {organization} {IOP Publishing},\
  \bibinfo {year} {2023})\ p.\ \bibinfo {pages} {012014}\BibitemShut {NoStop}%
\bibitem [{\citenamefont {Sakai}\ and\ \citenamefont
  {Arita}(2019)}]{sakai2019exotic}%
  \BibitemOpen
  \bibfield  {author} {\bibinfo {author} {\bibfnamefont {S.}~\bibnamefont
  {Sakai}}\ and\ \bibinfo {author} {\bibfnamefont {R.}~\bibnamefont {Arita}},\
  }\href@noop {} {\bibfield  {journal} {\bibinfo  {journal} {Physical Review
  Research}\ }\textbf {\bibinfo {volume} {1}},\ \bibinfo {pages} {022002}
  (\bibinfo {year} {2019})}\BibitemShut {NoStop}%
\bibitem [{\citenamefont {Ghadimi}\ \emph {et~al.}(2021)\citenamefont
  {Ghadimi}, \citenamefont {Sugimoto}, \citenamefont {Tanaka},\ and\
  \citenamefont {Tohyama}}]{ghadimi2021topological}%
  \BibitemOpen
  \bibfield  {author} {\bibinfo {author} {\bibfnamefont {R.}~\bibnamefont
  {Ghadimi}}, \bibinfo {author} {\bibfnamefont {T.}~\bibnamefont {Sugimoto}},
  \bibinfo {author} {\bibfnamefont {K.}~\bibnamefont {Tanaka}}, \ and\ \bibinfo
  {author} {\bibfnamefont {T.}~\bibnamefont {Tohyama}},\ }\href@noop {}
  {\bibfield  {journal} {\bibinfo  {journal} {Physical Review B}\ }\textbf
  {\bibinfo {volume} {104}},\ \bibinfo {pages} {144511} (\bibinfo {year}
  {2021})}\BibitemShut {NoStop}%
\bibitem [{\citenamefont {Cao}\ \emph {et~al.}(2020)\citenamefont {Cao},
  \citenamefont {Zhang}, \citenamefont {Liu}, \citenamefont {Liu},
  \citenamefont {Chen},\ and\ \citenamefont {Yang}}]{cao2020kohn}%
  \BibitemOpen
  \bibfield  {author} {\bibinfo {author} {\bibfnamefont {Y.}~\bibnamefont
  {Cao}}, \bibinfo {author} {\bibfnamefont {Y.}~\bibnamefont {Zhang}}, \bibinfo
  {author} {\bibfnamefont {Y.-B.}\ \bibnamefont {Liu}}, \bibinfo {author}
  {\bibfnamefont {C.-C.}\ \bibnamefont {Liu}}, \bibinfo {author} {\bibfnamefont
  {W.-Q.}\ \bibnamefont {Chen}}, \ and\ \bibinfo {author} {\bibfnamefont
  {F.}~\bibnamefont {Yang}},\ }\href@noop {} {\bibfield  {journal} {\bibinfo
  {journal} {Physical Review Letters}\ }\textbf {\bibinfo {volume} {125}},\
  \bibinfo {pages} {017002} (\bibinfo {year} {2020})}\BibitemShut {NoStop}%
\bibitem [{\citenamefont {Liu}\ \emph {et~al.}(2023{\natexlab{b}})\citenamefont
  {Liu}, \citenamefont {Zhang}, \citenamefont {Chen},\ and\ \citenamefont
  {Yang}}]{liu2023high}%
  \BibitemOpen
  \bibfield  {author} {\bibinfo {author} {\bibfnamefont {Y.-B.}\ \bibnamefont
  {Liu}}, \bibinfo {author} {\bibfnamefont {Y.}~\bibnamefont {Zhang}}, \bibinfo
  {author} {\bibfnamefont {W.-Q.}\ \bibnamefont {Chen}}, \ and\ \bibinfo
  {author} {\bibfnamefont {F.}~\bibnamefont {Yang}},\ }\href@noop {} {\bibfield
   {journal} {\bibinfo  {journal} {Physical Review B}\ }\textbf {\bibinfo
  {volume} {107}},\ \bibinfo {pages} {014501} (\bibinfo {year}
  {2023}{\natexlab{b}})}\BibitemShut {NoStop}%
\bibitem [{\citenamefont {Trabelsi}\ \emph
  {et~al.}(2019{\natexlab{a}})\citenamefont {Trabelsi}, \citenamefont
  {Ben~Ali}, \citenamefont {Belhadj},\ and\ \citenamefont
  {Kanzari}}]{trabelsi2019photonic}%
  \BibitemOpen
  \bibfield  {author} {\bibinfo {author} {\bibfnamefont {Y.}~\bibnamefont
  {Trabelsi}}, \bibinfo {author} {\bibfnamefont {N.}~\bibnamefont {Ben~Ali}},
  \bibinfo {author} {\bibfnamefont {W.}~\bibnamefont {Belhadj}}, \ and\
  \bibinfo {author} {\bibfnamefont {M.}~\bibnamefont {Kanzari}},\ }\href@noop
  {} {\bibfield  {journal} {\bibinfo  {journal} {Journal of Superconductivity
  and Novel Magnetism}\ }\textbf {\bibinfo {volume} {32}},\ \bibinfo {pages}
  {3541} (\bibinfo {year} {2019}{\natexlab{a}})}\BibitemShut {NoStop}%
\bibitem [{\citenamefont {Trabelsi}\ \emph
  {et~al.}(2019{\natexlab{b}})\citenamefont {Trabelsi}, \citenamefont {Ali},
  \citenamefont {Elhawil}, \citenamefont {Krishnamurthy}, \citenamefont
  {Kanzari}, \citenamefont {Amiri},\ and\ \citenamefont
  {Yupapin}}]{trabelsi2019design}%
  \BibitemOpen
  \bibfield  {author} {\bibinfo {author} {\bibfnamefont {Y.}~\bibnamefont
  {Trabelsi}}, \bibinfo {author} {\bibfnamefont {N.~B.}\ \bibnamefont {Ali}},
  \bibinfo {author} {\bibfnamefont {A.}~\bibnamefont {Elhawil}}, \bibinfo
  {author} {\bibfnamefont {R.}~\bibnamefont {Krishnamurthy}}, \bibinfo {author}
  {\bibfnamefont {M.}~\bibnamefont {Kanzari}}, \bibinfo {author} {\bibfnamefont
  {I.~S.}\ \bibnamefont {Amiri}}, \ and\ \bibinfo {author} {\bibfnamefont
  {P.}~\bibnamefont {Yupapin}},\ }\href@noop {} {\bibfield  {journal} {\bibinfo
   {journal} {Results in Physics}\ }\textbf {\bibinfo {volume} {13}},\ \bibinfo
  {pages} {102343} (\bibinfo {year} {2019}{\natexlab{b}})}\BibitemShut
  {NoStop}%
\bibitem [{\citenamefont {Wu}\ and\ \citenamefont
  {Gao}(2012)}]{wu2012transmission}%
  \BibitemOpen
  \bibfield  {author} {\bibinfo {author} {\bibfnamefont {J.-j.}\ \bibnamefont
  {Wu}}\ and\ \bibinfo {author} {\bibfnamefont {J.-x.}\ \bibnamefont {Gao}},\
  }\href@noop {} {\bibfield  {journal} {\bibinfo  {journal} {Optik}\ }\textbf
  {\bibinfo {volume} {123}},\ \bibinfo {pages} {986} (\bibinfo {year}
  {2012})}\BibitemShut {NoStop}%
\bibitem [{\citenamefont {Wo{\l}oszyn}\ and\ \citenamefont
  {Spisak}(2012)}]{woloszyn2012multifractal}%
  \BibitemOpen
  \bibfield  {author} {\bibinfo {author} {\bibfnamefont {M.}~\bibnamefont
  {Wo{\l}oszyn}}\ and\ \bibinfo {author} {\bibfnamefont {B.~J.}\ \bibnamefont
  {Spisak}},\ }\href@noop {} {\bibfield  {journal} {\bibinfo  {journal} {The
  European Physical Journal B}\ }\textbf {\bibinfo {volume} {85}},\ \bibinfo
  {pages} {1} (\bibinfo {year} {2012})}\BibitemShut {NoStop}%
\bibitem [{\citenamefont {Korol}\ and\ \citenamefont
  {Isai}(2013)}]{korol2013energy}%
  \BibitemOpen
  \bibfield  {author} {\bibinfo {author} {\bibfnamefont {A.}~\bibnamefont
  {Korol}}\ and\ \bibinfo {author} {\bibfnamefont {V.}~\bibnamefont {Isai}},\
  }\href@noop {} {\bibfield  {journal} {\bibinfo  {journal} {Physics of the
  solid state}\ }\textbf {\bibinfo {volume} {55}},\ \bibinfo {pages} {2596}
  (\bibinfo {year} {2013})}\BibitemShut {NoStop}%
\bibitem [{\citenamefont {Todd}\ \emph {et~al.}(1986)\citenamefont {Todd},
  \citenamefont {Merlin}, \citenamefont {Clarke}, \citenamefont {Mohanty},\
  and\ \citenamefont {Axe}}]{todd1986synchrotron}%
  \BibitemOpen
  \bibfield  {author} {\bibinfo {author} {\bibfnamefont {J.}~\bibnamefont
  {Todd}}, \bibinfo {author} {\bibfnamefont {R.}~\bibnamefont {Merlin}},
  \bibinfo {author} {\bibfnamefont {R.}~\bibnamefont {Clarke}}, \bibinfo
  {author} {\bibfnamefont {K.}~\bibnamefont {Mohanty}}, \ and\ \bibinfo
  {author} {\bibfnamefont {J.}~\bibnamefont {Axe}},\ }\href@noop {} {\bibfield
  {journal} {\bibinfo  {journal} {Physical review letters}\ }\textbf {\bibinfo
  {volume} {57}},\ \bibinfo {pages} {1157} (\bibinfo {year}
  {1986})}\BibitemShut {NoStop}%
\bibitem [{\citenamefont {Cohn}\ \emph {et~al.}(1988)\citenamefont {Cohn},
  \citenamefont {Lin}, \citenamefont {Lamelas}, \citenamefont {He},
  \citenamefont {Clarke},\ and\ \citenamefont {Uher}}]{cohn1988upper}%
  \BibitemOpen
  \bibfield  {author} {\bibinfo {author} {\bibfnamefont {J.}~\bibnamefont
  {Cohn}}, \bibinfo {author} {\bibfnamefont {J.-J.}\ \bibnamefont {Lin}},
  \bibinfo {author} {\bibfnamefont {F.}~\bibnamefont {Lamelas}}, \bibinfo
  {author} {\bibfnamefont {H.}~\bibnamefont {He}}, \bibinfo {author}
  {\bibfnamefont {R.}~\bibnamefont {Clarke}}, \ and\ \bibinfo {author}
  {\bibfnamefont {C.}~\bibnamefont {Uher}},\ }\href@noop {} {\bibfield
  {journal} {\bibinfo  {journal} {Physical Review B}\ }\textbf {\bibinfo
  {volume} {38}},\ \bibinfo {pages} {2326} (\bibinfo {year}
  {1988})}\BibitemShut {NoStop}%
\bibitem [{\citenamefont {Zhu}\ \emph {et~al.}(1997)\citenamefont {Zhu},
  \citenamefont {Zhu}, \citenamefont {Qin}, \citenamefont {Wang}, \citenamefont
  {Ge},\ and\ \citenamefont {Ming}}]{zhu1997experimental}%
  \BibitemOpen
  \bibfield  {author} {\bibinfo {author} {\bibfnamefont {S.-n.}\ \bibnamefont
  {Zhu}}, \bibinfo {author} {\bibfnamefont {Y.-y.}\ \bibnamefont {Zhu}},
  \bibinfo {author} {\bibfnamefont {Y.-q.}\ \bibnamefont {Qin}}, \bibinfo
  {author} {\bibfnamefont {H.-f.}\ \bibnamefont {Wang}}, \bibinfo {author}
  {\bibfnamefont {C.-z.}\ \bibnamefont {Ge}}, \ and\ \bibinfo {author}
  {\bibfnamefont {N.-b.}\ \bibnamefont {Ming}},\ }\href@noop {} {\bibfield
  {journal} {\bibinfo  {journal} {Physical review letters}\ }\textbf {\bibinfo
  {volume} {78}},\ \bibinfo {pages} {2752} (\bibinfo {year}
  {1997})}\BibitemShut {NoStop}%
\bibitem [{\citenamefont {Jagannathan}(2021)}]{jagannathan2021fibonacci}%
  \BibitemOpen
  \bibfield  {author} {\bibinfo {author} {\bibfnamefont {A.}~\bibnamefont
  {Jagannathan}},\ }\href@noop {} {\bibfield  {journal} {\bibinfo  {journal}
  {Reviews of Modern Physics}\ }\textbf {\bibinfo {volume} {93}},\ \bibinfo
  {pages} {045001} (\bibinfo {year} {2021})}\BibitemShut {NoStop}%
\bibitem [{\citenamefont {Sire}\ and\ \citenamefont
  {Mosseri}(1989)}]{sireSpectrum1DQuasicrystals1989}%
  \BibitemOpen
  \bibfield  {author} {\bibinfo {author} {\bibfnamefont {C.}~\bibnamefont
  {Sire}}\ and\ \bibinfo {author} {\bibfnamefont {R.}~\bibnamefont {Mosseri}},\
  }\href {\doibase 10.1051/jphys:0198900500240344700} {\bibfield  {journal}
  {\bibinfo  {journal} {Journal de Physique}\ }\textbf {\bibinfo {volume}
  {50}},\ \bibinfo {pages} {3447} (\bibinfo {year} {1989})}\BibitemShut
  {NoStop}%
\bibitem [{\citenamefont {You}\ and\ \citenamefont {Hu}(1988)}]{you1988global}%
  \BibitemOpen
  \bibfield  {author} {\bibinfo {author} {\bibfnamefont {J.}~\bibnamefont
  {You}}\ and\ \bibinfo {author} {\bibfnamefont {T.}~\bibnamefont {Hu}},\
  }\href@noop {} {\bibfield  {journal} {\bibinfo  {journal} {physica status
  solidi (b)}\ }\textbf {\bibinfo {volume} {147}},\ \bibinfo {pages} {471}
  (\bibinfo {year} {1988})}\BibitemShut {NoStop}%
\bibitem [{\citenamefont {Feingold}\ \emph {et~al.}(1988)\citenamefont
  {Feingold}, \citenamefont {Kadanoff},\ and\ \citenamefont
  {Piro}}]{feingold1988universalities}%
  \BibitemOpen
  \bibfield  {author} {\bibinfo {author} {\bibfnamefont {M.}~\bibnamefont
  {Feingold}}, \bibinfo {author} {\bibfnamefont {L.}~\bibnamefont {Kadanoff}},
  \ and\ \bibinfo {author} {\bibfnamefont {O.}~\bibnamefont {Piro}},\ }in\
  \href@noop {} {\emph {\bibinfo {booktitle} {Springer Proc. Phys.}}}\
  (\bibinfo {organization} {Springer},\ \bibinfo {year} {1988})\BibitemShut
  {NoStop}%
\bibitem [{\citenamefont {Sire}\ and\ \citenamefont
  {Mosseri}(1990)}]{sire1990excitation}%
  \BibitemOpen
  \bibfield  {author} {\bibinfo {author} {\bibfnamefont {C.}~\bibnamefont
  {Sire}}\ and\ \bibinfo {author} {\bibfnamefont {R.}~\bibnamefont {Mosseri}},\
  }\href@noop {} {\bibfield  {journal} {\bibinfo  {journal} {Journal de
  Physique}\ }\textbf {\bibinfo {volume} {51}},\ \bibinfo {pages} {1569}
  (\bibinfo {year} {1990})}\BibitemShut {NoStop}%
\bibitem [{\citenamefont {Rai}\ \emph {et~al.}(2021)\citenamefont {Rai},
  \citenamefont {Schl{\"o}mer}, \citenamefont {Matsumura}, \citenamefont
  {Haas},\ and\ \citenamefont
  {Jagannathan}}]{raiBulkTopologicalSignatures2021}%
  \BibitemOpen
  \bibfield  {author} {\bibinfo {author} {\bibfnamefont {G.}~\bibnamefont
  {Rai}}, \bibinfo {author} {\bibfnamefont {H.}~\bibnamefont {Schl{\"o}mer}},
  \bibinfo {author} {\bibfnamefont {C.}~\bibnamefont {Matsumura}}, \bibinfo
  {author} {\bibfnamefont {S.}~\bibnamefont {Haas}}, \ and\ \bibinfo {author}
  {\bibfnamefont {A.}~\bibnamefont {Jagannathan}},\ }\href {\doibase
  10.1103/PhysRevB.104.184202} {\bibfield  {journal} {\bibinfo  {journal}
  {Physical Review B}\ }\textbf {\bibinfo {volume} {104}},\ \bibinfo {pages}
  {184202} (\bibinfo {year} {2021})}\BibitemShut {NoStop}%
\bibitem [{\citenamefont {De~Gennes}\ and\ \citenamefont
  {Pincus}(2018)}]{de2018superconductivity}%
  \BibitemOpen
  \bibfield  {author} {\bibinfo {author} {\bibfnamefont {P.-G.}\ \bibnamefont
  {De~Gennes}}\ and\ \bibinfo {author} {\bibfnamefont {P.~A.}\ \bibnamefont
  {Pincus}},\ }\href@noop {} {\emph {\bibinfo {title} {Superconductivity of
  metals and alloys}}}\ (\bibinfo  {publisher} {CRC Press},\ \bibinfo {year}
  {2018})\BibitemShut {NoStop}%
\bibitem [{\citenamefont {Niu}\ and\ \citenamefont
  {Nori}(1990)}]{niu1990spectral}%
  \BibitemOpen
  \bibfield  {author} {\bibinfo {author} {\bibfnamefont {Q.}~\bibnamefont
  {Niu}}\ and\ \bibinfo {author} {\bibfnamefont {F.}~\bibnamefont {Nori}},\
  }\href@noop {} {\bibfield  {journal} {\bibinfo  {journal} {Physical Review
  B}\ }\textbf {\bibinfo {volume} {42}},\ \bibinfo {pages} {10329} (\bibinfo
  {year} {1990})}\BibitemShut {NoStop}%
\bibitem [{\citenamefont {Pi{\'e}chon}\ \emph {et~al.}(1995)\citenamefont
  {Pi{\'e}chon}, \citenamefont {Benakli},\ and\ \citenamefont
  {Jagannathan}}]{piechon1995analytical}%
  \BibitemOpen
  \bibfield  {author} {\bibinfo {author} {\bibfnamefont {F.}~\bibnamefont
  {Pi{\'e}chon}}, \bibinfo {author} {\bibfnamefont {M.}~\bibnamefont
  {Benakli}}, \ and\ \bibinfo {author} {\bibfnamefont {A.}~\bibnamefont
  {Jagannathan}},\ }\href@noop {} {\bibfield  {journal} {\bibinfo  {journal}
  {Physical review letters}\ }\textbf {\bibinfo {volume} {74}},\ \bibinfo
  {pages} {5248} (\bibinfo {year} {1995})}\BibitemShut {NoStop}%
\bibitem [{\citenamefont {Ghosal}\ \emph {et~al.}(2001)\citenamefont {Ghosal},
  \citenamefont {Randeria},\ and\ \citenamefont
  {Trivedi}}]{ghosalInhomogeneousPairingHighly2001}%
  \BibitemOpen
  \bibfield  {author} {\bibinfo {author} {\bibfnamefont {A.}~\bibnamefont
  {Ghosal}}, \bibinfo {author} {\bibfnamefont {M.}~\bibnamefont {Randeria}}, \
  and\ \bibinfo {author} {\bibfnamefont {N.}~\bibnamefont {Trivedi}},\ }\href
  {\doibase 10.1103/PhysRevB.65.014501} {\bibfield  {journal} {\bibinfo
  {journal} {Physical Review B}\ }\textbf {\bibinfo {volume} {65}},\ \bibinfo
  {pages} {014501} (\bibinfo {year} {2001})}\BibitemShut {NoStop}%
\bibitem [{\citenamefont {Ghosal}\ \emph {et~al.}(1998)\citenamefont {Ghosal},
  \citenamefont {Randeria},\ and\ \citenamefont
  {Trivedi}}]{ghosalRoleSpatialAmplitude1998}%
  \BibitemOpen
  \bibfield  {author} {\bibinfo {author} {\bibfnamefont {A.}~\bibnamefont
  {Ghosal}}, \bibinfo {author} {\bibfnamefont {M.}~\bibnamefont {Randeria}}, \
  and\ \bibinfo {author} {\bibfnamefont {N.}~\bibnamefont {Trivedi}},\ }\href
  {\doibase 10.1103/PhysRevLett.81.3940} {\bibfield  {journal} {\bibinfo
  {journal} {Physical Review Letters}\ }\textbf {\bibinfo {volume} {81}},\
  \bibinfo {pages} {3940} (\bibinfo {year} {1998})}\BibitemShut {NoStop}%
\bibitem [{\citenamefont {Fan}\ \emph {et~al.}(2021)\citenamefont {Fan},
  \citenamefont {Chern},\ and\ \citenamefont {Lin}}]{fan2021enhanced}%
  \BibitemOpen
  \bibfield  {author} {\bibinfo {author} {\bibfnamefont {Z.}~\bibnamefont
  {Fan}}, \bibinfo {author} {\bibfnamefont {G.-W.}\ \bibnamefont {Chern}}, \
  and\ \bibinfo {author} {\bibfnamefont {S.-Z.}\ \bibnamefont {Lin}},\
  }\href@noop {} {\bibfield  {journal} {\bibinfo  {journal} {Physical Review
  Research}\ }\textbf {\bibinfo {volume} {3}},\ \bibinfo {pages} {023195}
  (\bibinfo {year} {2021})}\BibitemShut {NoStop}%
\bibitem [{\citenamefont {Oliveira}\ \emph {et~al.}(2023)\citenamefont
  {Oliveira}, \citenamefont {Gon{\c{c}}alves}, \citenamefont {Ribeiro},
  \citenamefont {Castro},\ and\ \citenamefont
  {Amorim}}]{oliveira2023incommensurability}%
  \BibitemOpen
  \bibfield  {author} {\bibinfo {author} {\bibfnamefont {R.}~\bibnamefont
  {Oliveira}}, \bibinfo {author} {\bibfnamefont {M.}~\bibnamefont
  {Gon{\c{c}}alves}}, \bibinfo {author} {\bibfnamefont {P.}~\bibnamefont
  {Ribeiro}}, \bibinfo {author} {\bibfnamefont {E.~V.}\ \bibnamefont {Castro}},
  \ and\ \bibinfo {author} {\bibfnamefont {B.}~\bibnamefont {Amorim}},\
  }\href@noop {} {\bibfield  {journal} {\bibinfo  {journal} {arXiv preprint
  arXiv:2303.17656}\ } (\bibinfo {year} {2023})}\BibitemShut {NoStop}%
\bibitem [{\citenamefont {Noda}\ \emph
  {et~al.}(2015{\natexlab{a}})\citenamefont {Noda}, \citenamefont {Inaba},\
  and\ \citenamefont {Yamashita}}]{noda2015bcs}%
  \BibitemOpen
  \bibfield  {author} {\bibinfo {author} {\bibfnamefont {K.}~\bibnamefont
  {Noda}}, \bibinfo {author} {\bibfnamefont {K.}~\bibnamefont {Inaba}}, \ and\
  \bibinfo {author} {\bibfnamefont {M.}~\bibnamefont {Yamashita}},\ }\href@noop
  {} {\bibfield  {journal} {\bibinfo  {journal} {arXiv:1512.07858}\ } (\bibinfo
  {year} {2015}{\natexlab{a}})}\BibitemShut {NoStop}%
\bibitem [{\citenamefont {Noda}\ \emph
  {et~al.}(2015{\natexlab{b}})\citenamefont {Noda}, \citenamefont {Inaba},\
  and\ \citenamefont {Yamashita}}]{noda2015magnetism}%
  \BibitemOpen
  \bibfield  {author} {\bibinfo {author} {\bibfnamefont {K.}~\bibnamefont
  {Noda}}, \bibinfo {author} {\bibfnamefont {K.}~\bibnamefont {Inaba}}, \ and\
  \bibinfo {author} {\bibfnamefont {M.}~\bibnamefont {Yamashita}},\ }\href@noop
  {} {\bibfield  {journal} {\bibinfo  {journal} {Physical Review A}\ }\textbf
  {\bibinfo {volume} {91}},\ \bibinfo {pages} {063610} (\bibinfo {year}
  {2015}{\natexlab{b}})}\BibitemShut {NoStop}%
\bibitem [{\citenamefont {Kohmoto}\ \emph {et~al.}(1987)\citenamefont
  {Kohmoto}, \citenamefont {Sutherland},\ and\ \citenamefont
  {Tang}}]{kohmoto1987critical}%
  \BibitemOpen
  \bibfield  {author} {\bibinfo {author} {\bibfnamefont {M.}~\bibnamefont
  {Kohmoto}}, \bibinfo {author} {\bibfnamefont {B.}~\bibnamefont {Sutherland}},
  \ and\ \bibinfo {author} {\bibfnamefont {C.}~\bibnamefont {Tang}},\
  }\href@noop {} {\bibfield  {journal} {\bibinfo  {journal} {Physical Review
  B}\ }\textbf {\bibinfo {volume} {35}},\ \bibinfo {pages} {1020} (\bibinfo
  {year} {1987})}\BibitemShut {NoStop}%
\bibitem [{\citenamefont {Zhong}\ \emph {et~al.}(1995)\citenamefont {Zhong},
  \citenamefont {Bellissard},\ and\ \citenamefont {Mosseri}}]{zhong1995green}%
  \BibitemOpen
  \bibfield  {author} {\bibinfo {author} {\bibfnamefont {J.}~\bibnamefont
  {Zhong}}, \bibinfo {author} {\bibfnamefont {J.}~\bibnamefont {Bellissard}}, \
  and\ \bibinfo {author} {\bibfnamefont {R.}~\bibnamefont {Mosseri}},\
  }\href@noop {} {\bibfield  {journal} {\bibinfo  {journal} {Journal of
  Physics: Condensed Matter}\ }\textbf {\bibinfo {volume} {7}},\ \bibinfo
  {pages} {3507} (\bibinfo {year} {1995})}\BibitemShut {NoStop}%
\bibitem [{\citenamefont {Mac{\'e}}\ \emph {et~al.}(2016)\citenamefont
  {Mac{\'e}}, \citenamefont {Jagannathan},\ and\ \citenamefont
  {Pi{\'e}chon}}]{mace2016fractal}%
  \BibitemOpen
  \bibfield  {author} {\bibinfo {author} {\bibfnamefont {N.}~\bibnamefont
  {Mac{\'e}}}, \bibinfo {author} {\bibfnamefont {A.}~\bibnamefont
  {Jagannathan}}, \ and\ \bibinfo {author} {\bibfnamefont {F.}~\bibnamefont
  {Pi{\'e}chon}},\ }\href@noop {} {\bibfield  {journal} {\bibinfo  {journal}
  {Physical Review B}\ }\textbf {\bibinfo {volume} {93}},\ \bibinfo {pages}
  {205153} (\bibinfo {year} {2016})}\BibitemShut {NoStop}%
\bibitem [{\citenamefont {Nozieres}\ and\ \citenamefont
  {Pistolesi}(1999)}]{nozieresSemiconductorsSuperconductorsSimple1999}%
  \BibitemOpen
  \bibfield  {author} {\bibinfo {author} {\bibfnamefont {P.}~\bibnamefont
  {Nozieres}}\ and\ \bibinfo {author} {\bibfnamefont {F.}~\bibnamefont
  {Pistolesi}},\ }\href {\doibase 10.1007/s100510050897} {\bibfield  {journal}
  {\bibinfo  {journal} {The European Physical Journal B}\ }\textbf {\bibinfo
  {volume} {10}},\ \bibinfo {pages} {649} (\bibinfo {year} {1999})},\ \bibinfo
  {note} {arXiv: cond-mat/9902273}\BibitemShut {NoStop}%
\bibitem [{\citenamefont {Niroula}\ \emph {et~al.}(2020)\citenamefont
  {Niroula}, \citenamefont {Rai}, \citenamefont {Haas},\ and\ \citenamefont
  {Kettemann}}]{niroulaSpatialBCSBECCrossover2020}%
  \BibitemOpen
  \bibfield  {author} {\bibinfo {author} {\bibfnamefont {A.}~\bibnamefont
  {Niroula}}, \bibinfo {author} {\bibfnamefont {G.}~\bibnamefont {Rai}},
  \bibinfo {author} {\bibfnamefont {S.}~\bibnamefont {Haas}}, \ and\ \bibinfo
  {author} {\bibfnamefont {S.}~\bibnamefont {Kettemann}},\ }\href {\doibase
  10.1103/PhysRevB.101.094514} {\bibfield  {journal} {\bibinfo  {journal}
  {Physical Review B}\ }\textbf {\bibinfo {volume} {101}},\ \bibinfo {pages}
  {094514} (\bibinfo {year} {2020})}\BibitemShut {NoStop}%
\bibitem [{\citenamefont {Feigel'Man}\ \emph {et~al.}(2010)\citenamefont
  {Feigel'Man}, \citenamefont {Ioffe}, \citenamefont {Kravtsov},\ and\
  \citenamefont {Cuevas}}]{feigel2010fractal}%
  \BibitemOpen
  \bibfield  {author} {\bibinfo {author} {\bibfnamefont {M.}~\bibnamefont
  {Feigel'Man}}, \bibinfo {author} {\bibfnamefont {L.}~\bibnamefont {Ioffe}},
  \bibinfo {author} {\bibfnamefont {V.}~\bibnamefont {Kravtsov}}, \ and\
  \bibinfo {author} {\bibfnamefont {E.}~\bibnamefont {Cuevas}},\ }\href@noop {}
  {\bibfield  {journal} {\bibinfo  {journal} {Annals of Physics}\ }\textbf
  {\bibinfo {volume} {325}},\ \bibinfo {pages} {1390} (\bibinfo {year}
  {2010})}\BibitemShut {NoStop}%
\end{thebibliography}%

\section{3D extension of the Fibonacci chain}  \label{3D FC}
\begin{figure} 
    \includegraphics[width = 1\columnwidth]{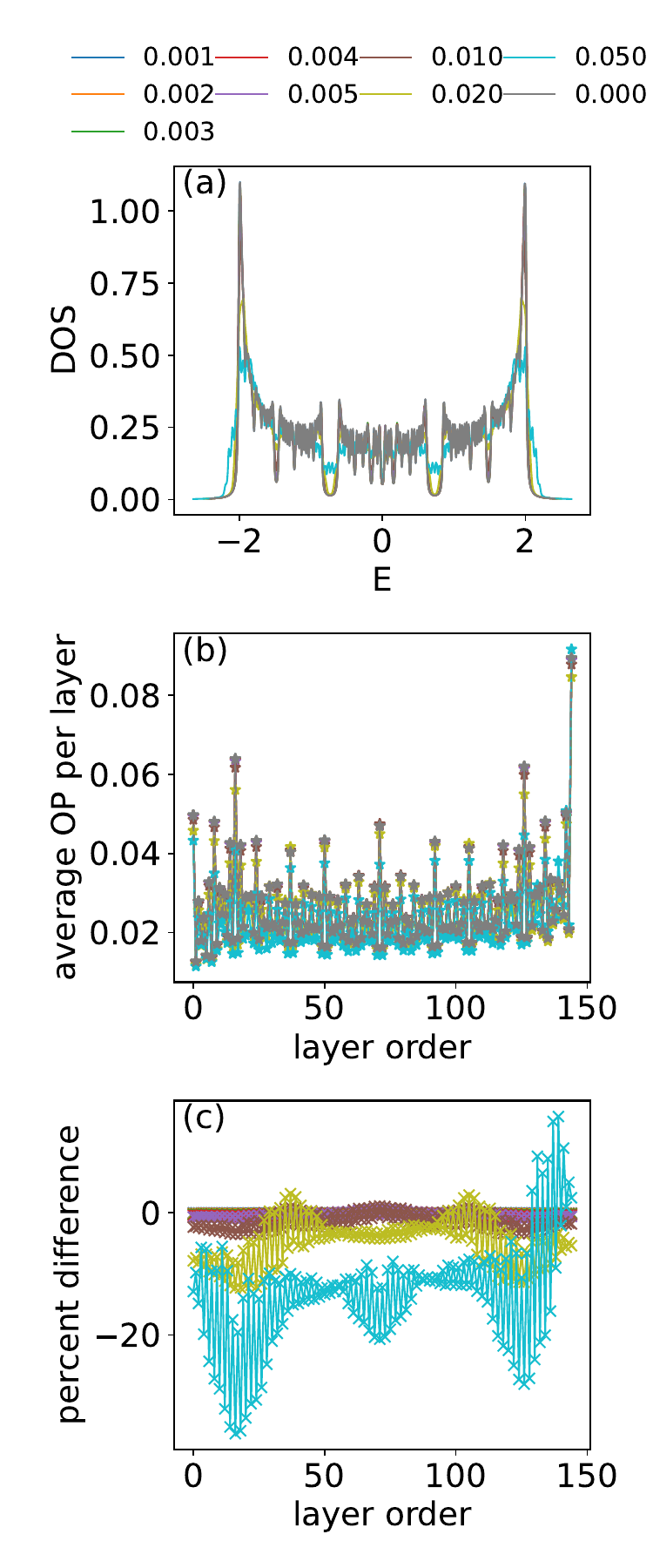}
    \caption{(a) Plots of density of states of Fibonacci superlattice with varing $t_p$. (b) Plots of the average OPs per layer of FC superlattice. (c) percent differences of average OPs relative to the 1D Fibonacci chain. The Fibonacci sequence length is 144. The modulation strength is $w=0.2$. The attraction strength is 1.0. These calculation parameters apply to all the calculations within this figure. Various $t_p$ values are denoted by the different colors indicated above panel (a).
}
    \label{FC superlattice}
\end{figure}

In order to verify our mean-field 1D simulation results are valid, we extend to the 3D Fibonacci superlattice, in which periodic 2D lattices are stacked along the third direction in a quasiperiodic way. The electronic spectrum and superconducting order parameters (OP) at half-filling are studied. The average hopping amplitude along the stacking direction is $t$. For simplicity, all the results below are reported in units of $t$. We tested the following in-plane hopping amplitudes $t_p$: 0.001, 0.002, 0.003, 0.004, 0.005, 0.01, 0.02, 0.05.  In Fig.~\ref{FC superlattice} (a), there is noticeable difference in the density of states (DOS) in FC superlattice when $t_p \geq 0.02$ compared with the 1D Fibonacci chain. The superconducting gapwidths are all the same no matter 3D superlattice or the 1D chain. When comparing the superconducting order parameter, the average OPs per layer are plotted against the single lattice OP of the 1D chain (Fig.~\ref{FC superlattice} (b)). Most of the average OPs are overlapped with 1D results. To see the differences better, the percent differences between the average OP per layer and OP of the 1D Fibonacci chain are shown in panel (c). There is a huge percent difference of order parameter as $t_p$ goes up to 0.02 and beyond. It's reasonable to simplify the 3D model to 1D when the in-plane interactions and properties are not dominating. Thus, when the in-plane hopping amplitudes are much less than those along the stacking direction, the mean-field results are valid applied to 1D systems.

\section{self-similar pattern in real and perpendicular space} \label{selfsimilarity}
\begin{figure*} 
    \includegraphics[width = 1\textwidth, trim = 20pt 15pt 10pt 300pt, clip]{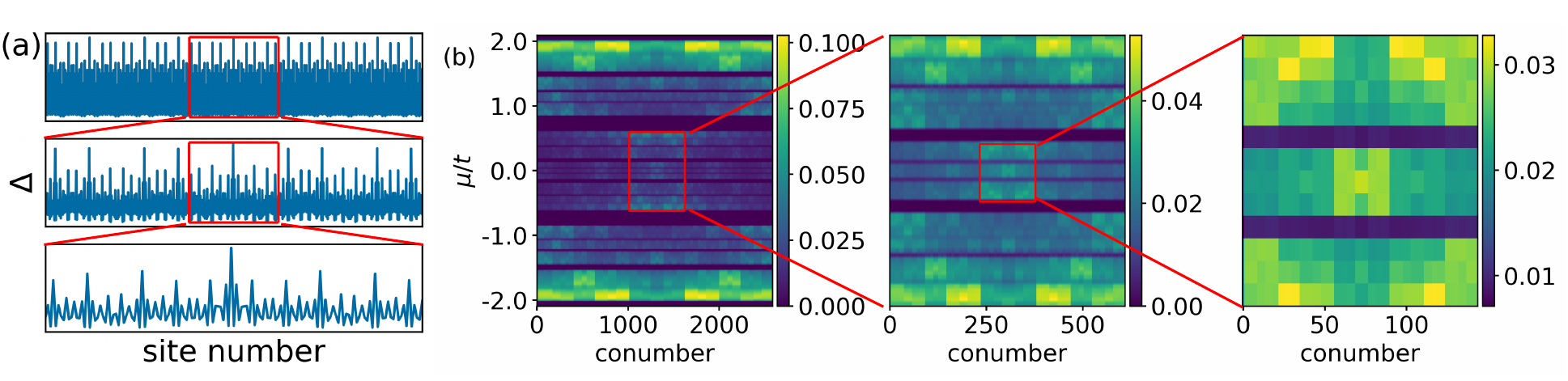}
    \caption{(a) superconducting order distribution in real space of a 2584-site chain at half-filling with $U=0.9t$ (top), central 610-site segment (middle), and central 144-site segment (bottom). (b) same figure as Fig.~\ref{fig:phasediagram} (a) (left), central 610-site plaquette (middle), central 144-site plaquette (right).
}
    \label{self-similar}
\end{figure*}

The self-similarity of superconducting order parameter distribution in real and perpendicular space is shown in Fig.~\ref{self-similar}. The central pattern (labeled by the red frame) is displayed in the next panel. Similar patterns recur scaled by the renormalization parameter $\tau^3$ \cite{niu1990spectral, mace2016fractal}. 

\section{Properties of the convergents of the golden ratio}
The golden ratio is the positive root of of the polynomial $x^2-x-1$. This implies two useful identities,
\begin{align}
    \tau^2 &= 1+\tau .\\
     \tau&=1 + \frac{1}{\tau}.
\end{align}
The analogous identities for the convergents are 
\begin{align}
    \tau_n\tau_{n-1} = 1 + \tau_{n-1},\\
    \tau_n = 1 + \frac{1}{\tau_{n-1}}.
\end{align}
The application of these identities allows us to compactly express the relation between the modulation strength $w$, the hoppings $t_A$ and $t_B$, and the hopping ratio $\rho = t_A/t_B$:
\begin{align}
    t_A &= 1 - \frac{w}{\tau_n\tau_{n-1}} \\&= 1 - \frac{w}{1+\tau_{n-1}},\\
    t_B &= 1 + \frac{w}{\tau_n} \\&= 1 + w(\tau_{n+1}-1),\\
    \rho &= 1 - \frac{w\tau_n}{\tau_n + w}.
\end{align}
For the  Fibonacci chain in the limit $n\to\infty$, the convergents $\tau_n$ can be replaced by the golden ratio $\tau$ in the above expressions.  When $\rho \to 0$, $\omega_{max} \approx 2.618$; when $\rho \to 1$, $\omega_{min} = 0$.

\section{Average order parameter as a function of temperature} \label{avgOPvsT}

\begin{figure}
    \includegraphics[width = 1\columnwidth, trim = 0pt 15pt 60pt 0pt, clip]{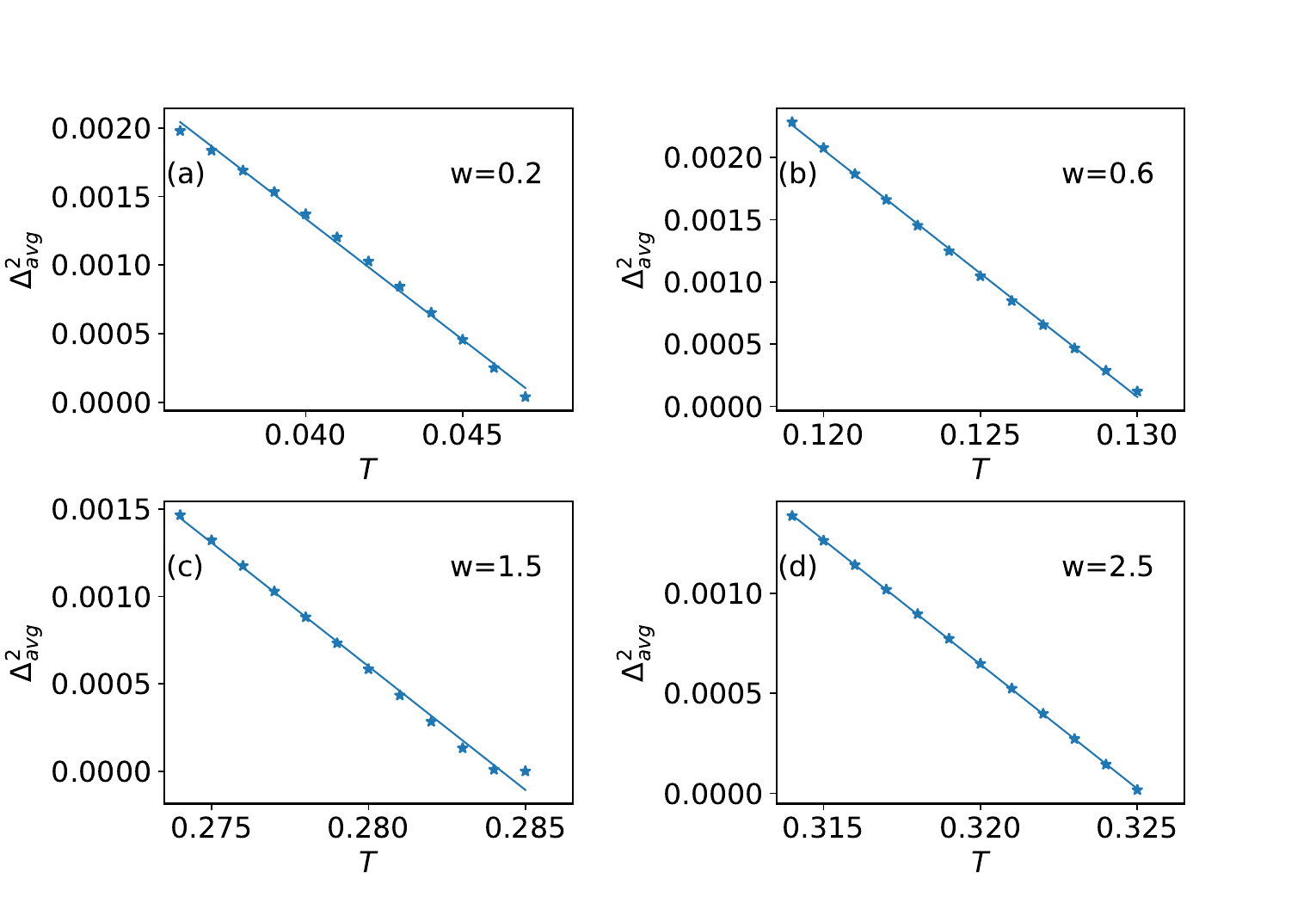}
    \caption{(a)-(d) The square of the average order parameter at different temperatures near its corresponding $T_c$ value for different modulation strengths. The dots represent the data points, and the curves are fits to  $\Delta^2_{avg} = const.\times (T_c - T)$ (which is the same  as $\Delta_{avg} = const.\times\sqrt{T_c - T}$). Here, the Fibonacci approximant length is 610, and the Hubbard attraction  is U=1.3t.
}
    \label{ameanOP-T}
\end{figure}

In Fig.~\ref{ameanOP-T}, the square of the average order parameter  is plotted as a function of temperature near the critical temperature. These curves follow the $\sqrt{T_c -T }$ relation as expected in mean field theory. The respective average order parameter magnitudes for atom sites and molecule sites also fit the same $\sqrt{T_c -T }$ relation just by multiplying different constant coefficients.

\section{Quadratic scaling of the superconducting gap for weak modulation strength}\label{app:sec:quadratic}
When the modulation strength vanishes, $w=0$, all eigenstates are doubly degenerate. $\epsilon_n^+ = \epsilon_n^- = \epsilon_n$. Turning on $w$, this degeneracy splits linearly in $w$ \cite{raiBulkTopologicalSignatures2021},
\begin{align}
    \epsilon_n^\pm = \epsilon_n \pm \alpha_n w,
\end{align}
where $2\alpha_n$ is the gap width of the pair of states at $\epsilon_n$. The contribution of a state at energy $\epsilon_n^\pm$ to the density of states at energy $\epsilon'$, $DOS(\epsilon')$, is a monotonically decreasing function of $\Delta \epsilon^\pm_n = |\epsilon_n^\pm - \epsilon'|$, which we denote as $f(\Delta \epsilon_n^\pm)$. Expanding $f(\Delta \epsilon_n^\pm)$ to second order in $w$,
\begin{align}
    DOS(\epsilon') =& \sum_n f(\Delta \epsilon_n^+) + f(\Delta \epsilon_n^-) \\
    =& \sum_n \left[2f(\Delta \epsilon_n) + f'(\Delta \epsilon_n)(\alpha_n w -\alpha_n w)\right.\nonumber\\
    &\left.+ f''(\Delta \epsilon_n)(\alpha _nw)^2\right] \\
    =& \sum_n 2f(\Delta \epsilon_n) + w^2 \sum_n f''(\Delta \epsilon_n) \alpha_n^2 \\
    =& DOS^{w=0}(\epsilon') + w^2 F(\epsilon'),
\end{align}
where we let $F(\epsilon')=\sum_n f''(\Delta \epsilon_n) \alpha_n^2$. Thus, in the regime of sufficiently small $w$, the density of states at a given energy $\epsilon'$ is proportional to the square of the modulation strength.

In BCS theory, $\Delta \propto e^{-\frac{1}{N(\epsilon_f)U}}$, where $N(\epsilon)$ is the density of states of the non-interacting system without any quasiperiodic modulation. We now Taylor expand this up to lowest order in $w^2$,
\begin{align}
    \Delta(w) \propto& e^{-\frac{1}{U\left(N(\epsilon_f)+w^2F(\epsilon_f)\right)}}\\
    =&e^{-\frac{1}{UN(\epsilon_f)}+\frac{w^2F(\epsilon_f)}{UN(\epsilon_f)^2} + \mathcal{O}(w^4)}\\
    =&e^{-\frac{1}{UN(\epsilon_f)}}\left(1+\frac{w^2F(\epsilon_f)}{UN(\epsilon_f)^2} + \mathcal{O}(w^4)
    \right)\\
    =& \Delta(\omega=0)\left(1+\frac{w^2\beta(\epsilon_f)}{UN(\epsilon_f)}\right)+ \mathcal{O}(w^4),
\end{align}
where $\beta(\epsilon_f) = F(\epsilon_f)/N(\epsilon_f)$.

\section{Critical temperature in the strong modulation limit} \label{Tc-larger-w}
The gap equation at different temperatures is
\begin{align}
    1 = U \int d\epsilon \frac{\rho(\epsilon)}{2E} \tanh\left(\frac{E}{2k_B T}\right),
\end{align}
where $E=\sqrt{\epsilon^2 + \Delta^2}$. When $w \to w_{max}$, the density of states near the Fermi level can be approximated by a delta function, and considering $\epsilon \to 0$ at the Fermi level, the gap equation can be simplified,
\begin{align}
    1 = \frac{U}{2\Delta}\tanh\left(\frac{\Delta}{2k_B T}\right).
    \label{eq: delta T pair}
\end{align}
Near the transition temperature $T_c$, $\Delta \to 0$. Then Eq. \ref{eq: delta T pair} can be written in the limiting form,
\begin{align}
    \frac{U}{2}\lim_{\Delta \to 0} \frac{\tanh\left(\frac{\Delta}{2k_B T_c}\right)}{\Delta} = 1
\end{align}
This limit can be calculated by L'Hôpital's rule, 
\begin{align}
    \frac{U}{4k_B T_c} \lim_{\Delta \to 0} \frac{1}{\cosh^2\left(\frac{\Delta}{2k_B T_c}\right)} &= 1 \\
    \frac{U}{4k_B T_c} &= 1
\end{align}
Thus, $k_B T_c = \frac{U}{4} = 0.325$. This result is pretty close to the numerically calculated critical temperature 0.326.

\section{Ratio of gap width and critical temperature, and its dependence on modulation strength} \label{ratio of gapwidth and Tc and scaling with w}
\begin{figure} 
    \includegraphics[width = 1\columnwidth]{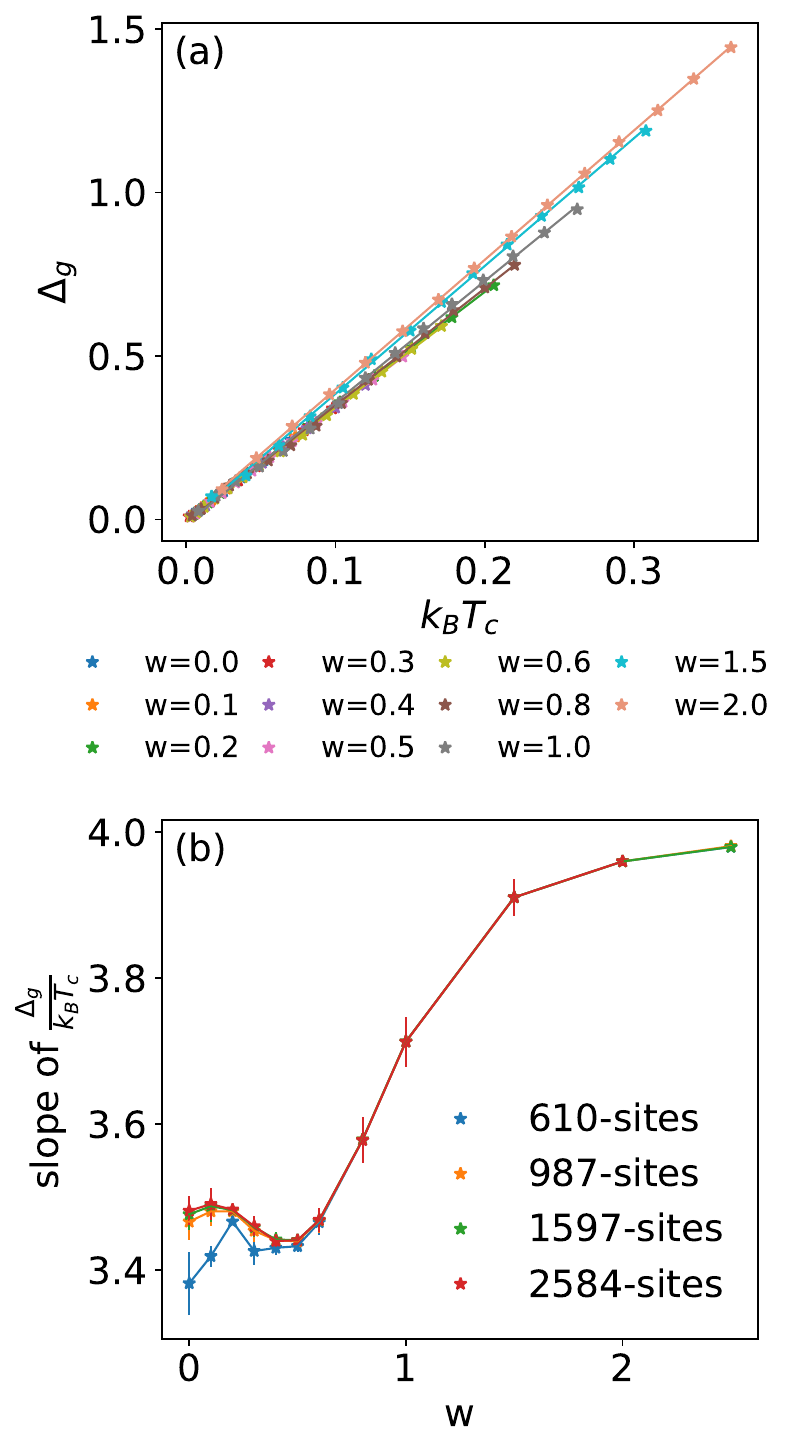}
    \caption{(a) Plots of gap width $\Delta_g$ vs. $k_B T_c$ of a 610-sites chain showing that they are proportional, with a $w$-dependent coefficient of proportionality. (b) Plots of the slopes of gap width $\Delta_g$ vs. $k_B T_c$ varing with modulation strength $w$ at Fibonacci chains with different lengths. As w increases from 0 to its maximum, this slope first increases and then drop to the minimum value near $w=0.5$, then increases to the largest value $\sim 4$.
}
    \label{gapwidthvsTc and STDEV}
\end{figure}

The coefficient of proportionality between $\Delta_g$ and $k_B T_c$ increases with $w$, as shown in Fig.~\ref{gapwidthvsTc and STDEV}(b).  We first analyze the slope of $\frac{\Delta_g}{k_B T_c}$ in the large $w$ limit. When $\rho\to 0$ (corresponding to $w\to $ $\sim$$2.6$), the chain is made up of disconnected \emph{atoms} (single sites connected by vanishing weak bonds) and \emph{molecules} (pairs of sites connected to each other by strong bonds and to the rest by weak bonds). In this limit, only the atom sites contribute to the pairing if the system is at half-filling, since their energy is zero and the Bogoliubov-de Gennes equation is represented by the matrix,
\begin{equation}
     H^{BdG}_{atom} =  
     \begin{pmatrix}
         0 & \Delta \\
         \Delta^{*} & 0
    \end{pmatrix}  ,
\end{equation}
 The positive-energy eigenvector of this matrix is $\frac{1}{\sqrt{2}}\begin{pmatrix}1\\1\end{pmatrix}$ regardless of the value of $\Delta$. Plugging this into $\Delta = U\sum_n v^*_{n}u_{n}$ where $n$ labels the positive-energy eigenvectors of the Bogoliubov-de Gennes pseudo-Hamiltonian. we find that the zero-temperature self-consistent order parameter is $U/2$. Therefore, the gap width is $\Delta_g \approx 2 \times \Delta_{atom} = U$. Also, $k_B T_c = \frac{U}{4}$ which has been proved in Appendix~\ref{Tc-larger-w}. Therefore, $\frac{\Delta_g}{k_B T_c} = 4$ in the large modulation strength limit, as seen in Fig.~\ref{gapwidthvsTc and STDEV}(b).

 In the small $w$ limit, the calculated slope value is 3.46, wich is close to that of BCS theory, $\frac{\Delta_g}{k_B T_c} = \frac{2\Delta}{k_B T_c} = 2 \times 1.76 = 3.52$.

\section{Intensity distribution of the local density of states in perpendicular  and energy space} \label{LDOSinconumber/energy}

In the non-interacting Fibonacci chain\cite{mace2016fractal},  the  spatial intensity pattern of the local density of states resembles that of the local superconducting order parameter shown in Fig.~\ref{fig:phasediagram} (a). Not surprisingly, the electron density distribution inherits the self-similar feature in perpendicular space. Comparison of Fig.~10 with Fig.~\ref{fig:phasediagram} (a) illustrates how the superconducting order parameter distribution follows the electron density distribution. 
\begin{figure}
    \includegraphics[width = 1\columnwidth]{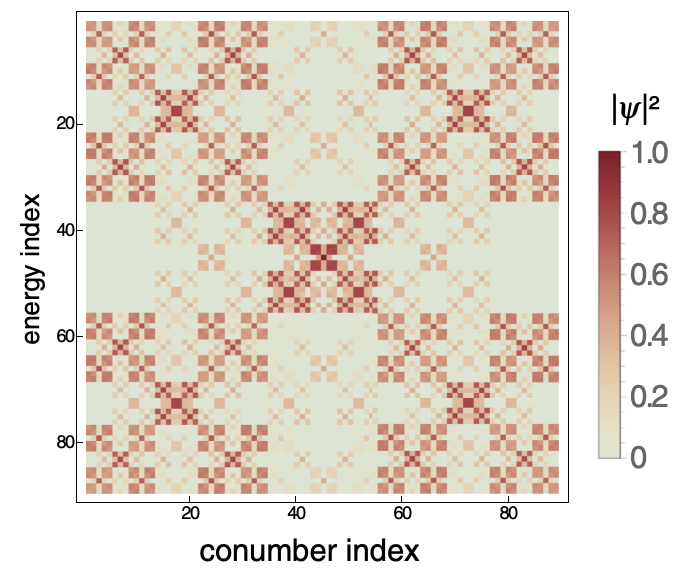}
    \caption{Non-interacting Fibonacci chain: spatial distribution of the LDOS intensity as a function of energy and conumber index. The darker colors represent  higher LDOS intensity. Reproduced with permission from \cite{mace2016fractal}. 
}
    \label{LDOS in conumber/energy}
\end{figure}

\end{document}